\documentstyle [preprint,epsfig,aps,12pt] {revtex}

\newcommand {\be} {\begin{equation}}
\newcommand {\bea} {\begin{eqnarray}}
\newcommand {\ee} {\end{equation}}
\newcommand {\eea} {\end{eqnarray}}
\newcommand {\bi} {\bibitem}
\newcommand {\r} {\vec{r}}
\newcommand {\x} {\vec{x}}
\newcommand {\z} {\vec{z}}
\newcommand {\y} {\vec{y}}
\newcommand {\p} {\psi}
\newcommand {\al} {\alpha}
\newcommand {\g} {\gamma}
\newcommand {\dd} {\tilde{\delta}}
\newcommand {\la} {\lambda}
\newcommand {\gd} {g^{\dagger}}
\newcommand {\fd} {f^{\dagger}}
\renewcommand{\theequation}{\thesection.\arabic{equation}}

\begin{document}
\title {Theory of continuum percolation III. Low density expansion}
\author{ Alon Drory$^1$, Brian Berkowitz$^2$, Giorgio Parisi$^1$, 
I.Balberg$^3$}
\address{\it  $^1$  Dipartimento di Fisica and INFN, Universit\`a 
{\sl La Sapienza}\\
   Piazzale Aldo Moro 2, Roma 00187, Italy \\
$^2$Department of Environmental Science and Energy Research, Weizmann 
Institute, Rehovot 76100, Israel\\
$^3$Department of Physics, Hebrew University, Jerusalem, Israel}
\maketitle

\begin{abstract}

We use a previously introduced mapping between the continuum percolation model
and the Potts fluid (a system of interacting  $s$-states spins which are 
free to move in the continuum) to derive the low density expansion of the 
pair connectedness and the mean cluster size. We prove that given an adequate 
identification of functions, the result is equivalent to the density expansion
derived from a completely different point of view by Coniglio {\it et al.} 
[J. Phys A {\bf 10}, 1123 (1977)] to describe physical clustering in a gas. We
then apply our expansion to a system of hypercubes with a hard core 
interaction. The calculated critical density is within approximately 5\% of the
results of simulations, and is thus much more precise than previous theoretical
results which were based on integral equations. We suggest that this is because
integral equations smooth out overly the partition function (i.e., they 
describe predominantly its analytical part), while our method targets instead 
the part which describes the phase transition (i.e., the singular part).
\end{abstract}


\renewcommand{\thesection}{\Roman{section}}
\section{INTRODUCTION}
\renewcommand{\thesection}{\arabic{section}}
\setcounter{equation}{0}

Two previous papers in this series \cite{drory}, hereafter referred to as I 
and II, presented a general formalism of continuum percolation \cite{rev}, 
where the system consists of classical particles interacting through a pair 
potential $ v(\r_i, \r_j)$, such that they can also bind (or connect) to each 
other with a probability $ p(\r_i,\r_j)$. Such a model is useful to describe
microemulsions \cite{bug}, composite materials \cite{mat} or some properties 
of water \cite{water}. In this model, the clustering depends on the density
$\rho$ of the particles. As the density increases, so does the mean cluster 
size $S$. At a well defined  critical density $\rho_c$, $S$ diverges, which 
signals the appearance of an infinite cluster. This is the percolation phase 
transition. Its critical behavior (i.e., the critical exponents) seem to be 
identical to the behavior of lattice percolation (see however a recent work by 
Okazaki {et al.} \cite{okazaki} which claims differently) but the percolation 
threshold -- the critical density -- is sensitive to all the details of the 
system.

The early theoretical attempts to calculate the percolation threshold 
culminated with the introduction by Balberg {\it et al.} of the notion of a 
critical 
total excluded volume $B_c$ \cite{bc}. The excluded volume is the volume around
one particle of the system in which the center of a second particle must be in 
order for the two particles to be connected. The total excluded volume of the
system is therefore another measure of the number of particles in the system, 
i.e., of the density. When measured in this way, the numerical value of the 
percolation threshold seemed to be relatively insensitive to the shape of the 
particles (unlike the density itself). It was therefore considered an 
approximately universal quantity. However, this concept could not take into 
account other properties of the system, such as interactions between the 
particles, nor could it explain the remaining dependence of the percolation 
threshold on the details of the system.

In 1977, Coniglio {\it et al.} proposed a different approach, based on a 
density expansion of the mean cluster size \cite{coniglio}. This in turn served
as the basis of an approximate calculation based on some integral equations 
analogous to those encountered in the theory of liquids \cite{hansen}. This 
approach yielded finally a theoretical prediction of the percolation threshold,
but the results remained only qualitative, discrepancies of up to 40 \% with 
computer simulations being common \cite{des,oth}.

Recently, however, Alon {\it et al.} \cite{alon} and Drory {\it et al.} 
\cite{dro} obtained quantitatively adequate results from Coniglio's expansion,
by using the density expansion directly instead of integral equations. This 
posed a curious problem, because the critical density is not low enough to 
suggest that a power  expansion should work. On the other hand, the expansion 
of Coniglio {\it et al.} had been developed to describe physical clustering in 
a gas, and its extension to general continuum percolation models was based on 
extensive analogies. It seemed that a more solid theoretical foundation was 
needed for continuum  percolation before this puzzle could be addressed. Such a
foundation has been laid in I and II, and we can now treat the problem of 
density expansions from a fresh point of view. The formalism presented in I 
and II is based on a quantitative mapping between the percolation model and an
extension of the Potts model, the Potts fluid. For easy reference, we recall 
here the essential definitions and results.

The $s$-state Potts fluid is a system of $N$ classical spins 
$\{\la_i\}_{i=1}^N$ interacting with each other through a spin-dependent pair 
potential $V(\r_i,\la_i;\r_j,\la_j)$, such that
\be
V(\r_i,\la_i;\r_j,\la_j) \equiv V(i,j) = \left\{ \begin{array}{r @{\quad 
     \mbox{if} \quad}l}
   U(\r_i,\r_j) & \la_i =\la_j \\ W(\r_i,\r_j) & \la_i \neq\la_j  .
   \end{array} \right.    \label {poten}
\ee
The spins are coupled to an external field $h(\r)$ through an interaction 
Hamiltonian 
\be
H_{int} = - \sum_{i=1}^N \,\p (\la_i) h(\r_i) ,
\ee
where
\be
\p (\la)  =  \left\{ \begin{array}{r @{\quad \mbox{if} \quad} l}
            s - 1 & \la = 1 \\ -1 & \la \ne 1 .\end{array} \right. \label{psi}
\ee

Up to some unimportant constants, the Potts fluid partition function (more
precisely, the configuration integral) is
\be
Z = \frac{1}{N!} \sum_{\{\la_m\}} \int \, d\r_1 \cdots d\r_N \,
\exp \left[ -\beta \sum_{i>j} V(i,j) + \beta \sum_{i=1}^N h(i) \p (\la_i) 
\right] .\label{conf}
\ee
The magnetization of the Potts fluid is defined as 
\be
M = \frac{1}{\beta N (s-1)}\, \frac{\partial \ln Z}{\partial h},
\label{defm}
\ee
where $h$ is the now constant external field. The susceptibility is
\be
\chi = \frac{\partial M}{\partial h}.
\label{defchi}
\ee

The $n$-density functions of the Potts fluid are defined as
\be
\rho^{(n)} (\r_1, \la_1; \ldots; \r_n, \la_n) = \frac{1}{Z(N-n)!} \int 
d\r_{n+1} \cdots d\r_N \exp\left[ - \beta \sum_{i>j} V(i,j) - \beta 
\sum_{i=1}^N h(i) \p (\la_i) \right] .\label{defr}
\ee
Of particular interest is the spin pair-distribution function, 
$g^{(2)}(\x,\al; \y, \g)$, 
defined as
\be
g^{(2)} (\x,\al; \y, \g) \equiv 
\frac{1}{\rho(\x) \rho(\y)}\, \rho^{(2)}(\x,\al;\y,\g) ,
\label{defg}
\ee
which tends to $1$ when $| \x - \y| \to \infty $. Here $\rho(\x)$ is the local
numerical density. It is often useful to define a spin correlation function 
$h^{(2)}(\x,\al; \y, \g)$ as
\be
h^{(2)}(\x,\al; \y, \g) \equiv g_s^{(2)} (\x,\al; \y, \g) - 1 .
\label{defh}
\ee
This function tends to zero when $| \x - \y| \to \infty $.

Any continuum percolation model defined by $v(i,j)$ and $p(i,j)$ can be 
mapped onto an appropriate Potts fluid model with a pair-spin interaction 
defined by
\bea
U(i,j) & = & v(i,j) , \nonumber\\
\exp \left[- \beta W(i,j) \right] & = & q(i,j) \exp \left[ - \beta v(i,j)
\right] ,
\label{map}
\eea
where 
\be
q(\r_i,\r_j) \equiv 1 - p(\r_i,\r_j)
\ee
is the probability of disconnection .

The relation between the Potts magnetization and the percolation probability,
$P(\rho)$, is 
\be
\lim_{h \to 0} \, \lim_{N \to \infty} \, \lim_{s \to 1} \, M = P(\rho) .
\label{prho}
\ee

For densities lower than the critical density $\rho_c$, the susceptibility is 
directly related to the mean cluster size, $S$.
\be
\lim_{h \to 0} \,\lim_{N \to \infty} \, \lim_{s \to 1} \, \chi = \beta S
\qquad\qquad (\rho < \rho_c) .
\label{chis}
\ee

An important quantity in the percolation model is the pair connectedness 
function $\gd (\x, \y)$, the meaning of which is
\bea
\rho(\x)\rho(\y) \, \gd (\x, \y)\,\, d \x \, d \y &=& 
\mbox{Probability of finding two particles in regions } \nonumber \\& & 
d \x \mbox{ and } d \y \mbox{ around  the positions }\x \mbox{ and } \y  
\mbox{, such} \nonumber \\
& &\mbox{that they both belong to the same cluster.} 
\eea
This function is related to the mean cluster size by
\be
S = 1 + \rho\int d \r \, \gd (\r) ,
\label{S}
\ee
where we assume, as we usually shall, that the system is translationally 
invariant, so that $\gd (\x, \y) = \gd (\x- \y)$ and 
$\rho(\x)= \rho(\y) = \rho$. 

The pair connectedness is related to the Potts pair correlation functions by
\bea
\gd (\x, \y) &=& \lim_{s \to 1} \left[ g_s^{(2)}( \x,\sigma ;
\y, \sigma) - g_s^{(2)}(\x,\sigma;\y, \eta) \right] \nonumber\\
&=& \lim_{s \to 1} \left[ h_s^{(2)}( \x,\sigma ;\y, \sigma) - 
h_s^{(2)}(\x,\sigma;\y, \eta) \right] .
\label{gdag}
\eea
where the spins $\sigma$ and $\eta$ are arbitrary except for the conditions 
$\sigma, \eta \ne 1$ and $\sigma \ne \eta$.

Recently, Drory has applied this formalism to a non-trivial one dimensional
model and managed to obtain its exact solution \cite{oned}. Continuing the
investigation of the usefulness of this formalism, we consider in the present 
work the region $\rho < \rho_c$
and derive series expansion in powers of the density for the mean cluster size
and the pair connectedness (the percolation probability, on the other hand, 
vanishes identically for these densities). To do this we introduce in section 
II the spin-functional differentiation, which is a simple generalization of 
the usual functional differentiation. With this tool, we obtain easily a 
diagrammatic expansion of the relevant quantities in section III. Section IV 
compares this expansion to the one derived by Coniglio {\it et al.} from a 
completely different starting point \cite{coniglio}. Section V then applies the
general results to a specific case, the extended hypercube models. The results
are presented in section VI. Finally in section VII, we discuss the reasons 
for the quantitative success of the present approach, compared with the 
inadequacy of previous analytical attempts.

\renewcommand{\thesection}{\Roman{section}}
\section{SPIN-FUNCTIONAL DIFFERENTIATION}
\renewcommand{\thesection}{\arabic{section}}
\setcounter{equation}{0}

The mathematical properties of the Potts fluid are very similar to the
corresponding ones for a classical fluid. In the following derivations, we
follow therefore very closely the presentation of Hansen and McDonald 
of the theory of classical fluids \cite{hansen}. Occasionally we shall skip 
some mathematical details which are identical for the Potts fluid and for the 
classical one.

Since all the quantities in which we are interested are expressible as 
statistical averages, we may choose to work in the grand canonical ensemble 
rather than in the canonical one.

The grand canonical partition function, $\Xi$, is given by
\be
\Xi = \sum_{N=0}^{\infty} \frac{1}{N!} \int d 1 \cdots d N 
\sum_{\la_1, \ldots, \la_N} \prod_{i=1}^N z^* (i, \la_i) \prod_{i>j}^N 
\exp \left[ - \beta V(i,j) \right] ,
\label{grand}
\ee
where
\be
z^* (i, \la_i) = \left( \frac{2 \pi \beta \hbar^2}{m} \right)^{3/2} \exp\left[
\beta \mu + \beta h(i) \p (\la_i) \right]
\label{act}
\ee
is the generalized activity. Here, $m$ is the mass of the particles 
(the spins), $\mu$ is the chemical potential, and $\hbar$ is Planck's constant.

The $n$-density functions are now defined to be
\be
\rho^{(n)} (\r_1, \la_1;\ldots; \r_n, \la_n) = \frac{1}{\Xi}\sum_{N=0}^{\infty}
\frac{1}{(N-n)!} \int d \r_{n+1} \cdots d \r_{N} \!\!\! \sum_{\la_{n+1}, 
\ldots, \la_N} \prod_{i=1}^N z^* (i, \la_i) \prod_{j>i}^N \exp \left[ - \beta 
V(i,j) \right] .\label{rdef}
\ee

Paper II presented a generalization of the functional derivative, which will
be called hereafter the spin-functional derivative. Let ${\cal{F}}[t(\r, \la)]$
be a functional of the function $t(\r, \la)$, which depends on a position 
variable  $\r$ as well as on an associated discrete spin variable $\la$. Then 
the spin-functional derivative $\dd {\cal{F}} / \delta t$ is defined through 
the relation
\be
\delta {\cal{F}}=\int d \r \,\sum_{\la} \frac{\dd {\cal{F}}}{\delta t(\r,\la)}
\, \delta t(\r, \la) ,
\label{deffd}
\ee
where $\delta {\cal{F}}$ is the change in ${\cal{F}}$ associated with a 
variation $\delta t$ in $ t(\r, \la)$. The only difference with the usual 
functional derivative is in the added summation over the spin variable. It is 
easily seen that this does not change any of the basic properties of the 
functional derivative operator. In particular, we have (see, e.g.,  Hansen and 
McDonald \cite{hansen}) that
\be
\frac{\dd t(\x, \al)}{\delta t(\y,\g)} = \delta (\x - \y) \, \delta_{\al, \g} ,
\label{delta}
\ee
and
\be
\frac{\dd}{\delta t(\y,\g)} \left[\int d \z \sum_{\la} t(\z, \la)\right] =
t(\y, \g) .  \label{prop1}
\ee
The equivalent of the change of variable formula is now
\be
\frac{\dd {\cal{F}}}{\delta u(\x,\al)} = \int d \y \sum_{\la} 
\frac{\dd {\cal{F}}}{\delta v(\x,\la)} \, \frac{\dd v(\x,\la)}
{\delta u(\x,\al)} .
\label{chvar}
\ee

By a generalization of Eq.~(\ref{prop1}), we now have that
\be
\frac{\dd \Xi}{\delta z^* (\x, \al)}
= \sum_{N=1}^{\infty} \frac{N}{N!} 
\int d 2 \cdots d N \sum_{\la_2, \ldots, \la_N} \prod_{i=2}^N 
z^* (i, \la_i) \prod_{i>j}^N \exp \left[ - \beta V(i,j) \right] ,
\label{delxi}
\ee
where $\r_1 =\x $ and $ \la_1 = \al $. Therefore, comparing with the definition
of the $n$-density functions, Eq.~(\ref{rdef}), we have that
\be
\rho^{(1)} (\x, \al) = \frac{z^* (\x, \al)}{\Xi} \frac{\dd \Xi}{\delta 
z^* (\x, \al)} = z^* (\x, \al) \frac{\dd \ln \Xi}{\delta z^* (\x, \al)} .
\label{rho1}
\ee
An immediate generalization yields
\be
\rho^{(n)} (1, \al_1; \ldots ; n, \al_n) = \frac{1}{\Xi}z^* (1,\al_1) \cdots 
z^* (n,\al_n) \frac{\dd^n \Xi}{\delta z^* (1,\al_1) \cdots \delta z^* 
(n,\al_n)} . \label{rhon}
\ee
In particular, combining Eqs.~(\ref{rho1}) and (\ref{rhon}), we have that
\be
\rho^{(2)} (\x, \al; \y, \g) - \rho^{(1)} (\x, \al)\rho^{(1)} (\y, \g) = 
z^* (\x, \al)z^* (\y, \g) \frac{\dd \ln \Xi}{\delta z^* (\x, \al)
\delta z^* (\y, \g)} .
\label{rho2}
\ee

A useful identity is obtained by using Eq.~(\ref{delta}) in conjunction with
Eq.~(\ref{rho1}),
\bea
\frac{\dd \rho^{(1)}(\x, \al)}{\delta \ln [z^* (\y, \g)]} &=& 
z^* (\y, \g)\frac{\dd}{\delta z^*(\y, \g)} \left[ z^*(\x, \al)\frac{\dd\ln \Xi}
{\delta z^*(\x, \al)}\right] \nonumber \\
&=& \rho^{(1)}(\x, \al) \delta (\x - \y) \,\delta_{\al, \g} + 
\rho^{(1)}(\x, \al)\rho^{(1)}(\y, \g)h^{(2)}(\x,\al;\y,\g) ,
\label{idrho}
\eea
where we have used Eqs.~(\ref{rho2}), (\ref{defg}) and (\ref{defh}).

By analogy with fluid systems \cite{hansen}, let us now define the spin direct
correlation function, $c(\x, \al; \y, \g)$, as
\be
c(\x, \al; \y, \g) \equiv \frac{\dd \ln \left[ \rho^{(1)}(\x, \al)
/ z^* (\x, \al) \right]}{\delta  \rho^{(1)}(\y, \g)} .
\label{defc}
\ee
Then, with the identity Eq.~(\ref{delta}), we have that
\be
\frac{\dd \ln [ z^* (\x, \al)]}{\delta  \rho^{(1)}(\y, \g)} = \frac{1}
{\rho^{(1)}(\x, \al)} \delta (\x - \y) \,\delta_{\al, \g} - c(\x, \al; \y,\g).
\label{idc}
\ee
Finally, from the change of variable formula, Eq.~(\ref{chvar}), we have that
\be
\delta (\x - \y) \,\delta_{\al, \g} = \frac{\dd \ln [ z^* (\x, \al)]}
{\delta \ln [ z^*(\y, \g)]} = \int d \z \sum_{\la} 
\frac{\dd \ln [ z^* (\x, \al)]}{\delta  \rho^{(1)}(\z, \la)}
\frac{\dd \rho^{(1)}(\z, \la)}{\delta \ln [ z^*(\y, \g)]} .
\label{idelta}
\ee
Substituting Eqs.~(\ref{idrho}) and (\ref{defc}) into Eq.~(\ref{idelta}), we
find
\be
h^{(2)}(\x, \al; \y, \g) = c(\x, \al; \y, \g) + \int d \z \sum_{\la} 
\rho^{(1)} (\z, \la) \, c(\x, \al; \z, \la) \, h^{(2)}(\z, \la; \y, \g) ,
\label{oz}
\ee
which is the equivalent, for the Potts fluid, of the classical Ornstein-Zernike
relation (see, e.g., Hansen and McDonald \cite{hansen}).

To find out the implications of this for the percolation system, let us 
substitute Eq.~(\ref{oz}) into the definition of the pair connectedness, 
Eq.~(\ref{defh}). Then, we have that
\bea
\lefteqn{h^{(2)}(\x, \al; \y, \al) - h^{(2)}(\x, \al; \y, \g)} \nonumber \\
&=& c(\x, \al; \y, \al) + \int d \z \, \rho^{(1)} (\z, \al) \, 
c(\x, \al; \z, \al)\, h^{(2)}(\z, \al; \y, \al) \nonumber \\
&+& \int d \z \sum_{\la \ne \al} \rho^{(1)} (\z, \la) \, c(\x, \al; \z, \la)\, 
h^{(2)}(\z, \la; \y, \al) - c(\x, \al; \y, \g) 
\nonumber \\
&-& \int d \z \, \rho^{(1)} (\z, \al) \, c(\x, \al; \z, \al)
\, h^{(2)}(\z, \al; \y, \g) 
- \int d \z \, \rho^{(1)} (\z, \g) \, c(\x, \al; \z, \g)\, 
h^{(2)}(\z, \g; \y, \g) \nonumber \\
&-& \int d \z \sum_{{\la \ne \al \atop \la \ne \g}} \rho^{(1)} (\z, \la) \, 
c(\x, \al; \z, \la) \, h^{(2)}(\z, \la; \y, \g) . \label{oz1}
\eea
We use now the fact that for $\rho< \rho_c$ the symmetry of the system is 
unbroken and therefore,
\be
\rho^{(1)} (\z, \la) = \frac{1}{s} \, \rho(\z) ,
\label{rhohom}
\ee
where $\rho(\z)$ is the local density at $\z$. For the same reason, we also 
have that
\bea
c(\x, \al; \y, \al) &=& c(\x, \g; \y, \g) , \nonumber\\
h^{(2)}(\x, \al; \y, \al) &=& h^{(2)}(\x, \g; \y, \g) , \nonumber \\
h^{(2)}(\x, \al; \y, \g) &=& h^{(2)}(\x, \al; \y, \la) ,
\eea
for any spins $\g, \la \ne \al$, and other similar relations. Then we can 
rewrite Eq.~(\ref{oz1}) as
\bea
\lefteqn{h^{(2)}(\x, \al; \y, \al) - h^{(2)}(\x, \al; \y, \g)} \nonumber \\
&=& c(\x, \al; \y, \al)  - c(\x, \al; \y, \g)
+ \frac{1}{s}\int d \z \, \rho(\z) \, c(\x, \al; \z, \al)\, 
h^{(2)}(\z, \al; \y, \al) \nonumber \\
&+& \frac{s-1}{s}\int d \z \, \rho (\z) \, c(\x, \al; \z, \g)\, 
h^{(2)}(\z, \g; \y, \g) 
- \frac{1}{s}\int d \z \, \rho (\z) \, c(\x, \al; \z, \al)
\, h^{(2)}(\z, \al; \y, \g) \nonumber \\
&-& \frac{1}{s}\int d \z \, \rho(\z) \, c(\x, \al; \z, \g)\, 
h^{(2)}(\z, \al; \y, \al) 
- \frac{s-2}{s}\int d \z \, \rho(\z) \, c(\x, \al; \z, \g) \, 
h^{(2)}(\z, \g; \y, \g) .
\label{oz2}
\eea
Finally, we can take the limit $s \to 1$. Equation (\ref{oz2}) then becomes
\bea
\lefteqn{\lim_{s \to 1} \, \left[ h^{(2)}(\x, \al; \y, \al) - h^{(2)}
(\x, \al; \y, \g)\right]} \nonumber \\
&=& \lim_{s \to 1} \,\Bigg \{ \left[ c(\x, \al; \y, \al)  - c(\x, \al; \y, \g)
\right] \nonumber \\
&+& \!\!\int d \z \,\rho(\z) \,\Big[ c(\x, \al; \z, \al) - c(\x, \al; \z, \g) 
\Big] \Big[ h^{(2)}(\z, \al; \y, \al) - h^{(2)}(\z, \al; \y, \g) \Big] 
\Bigg\} . \label{oz3}
\eea
We can now define a percolation direct connectedness function, 
$c^{\dagger}(\x, \y )$, as
\be
c^{\dagger}(\x, \y ) = \lim_{s \to 1} \,\left[ c(\x, \al; \y, \al)  - 
c(\x, \al; \y, \g)\right] .
\label{cdag}
\ee
Equation (\ref{oz3}) then becomes, with the help of the definition of the pair 
connectedness, Eq.~(\ref{gdag}),
\be
\gd (\x, \y) = c^{\dagger}(\x, \y) + \int d \z \, \rho(\z) \, c^{\dagger}
(\x,\z) \, \gd (\z,\y) ,
\label{ozper}
\ee
which is the percolation analog of the Ornstein-Zernike relation. This relation
will prove useful in the next section.

\renewcommand{\thesection}{\Roman{section}}
\section{ DIAGRAMMATIC EXPANSION}
\renewcommand{\thesection}{\arabic{section}}
\setcounter{equation}{0}

The density expansion of the pair correlation of a Potts fluid is obtained by 
following the steps leading to the corresponding density expansion for a 
classical fluid, once some simple generalizations have been made. The following
follows closely, therefore, the derivation presented in Hansen and McDonald 
\cite{hansen}.

First, let us define the Potts equivalent of the Mayer $f$-function, as
\be
\phi (\r_i, \la_i; \r_j, \la_j) \equiv \phi (i,j) = \exp\left[ - \beta 
V(\r_i, \la_i; \r_j, \la_j) \right] - 1 .
\label{defphi}
\ee
Let us now denote
\bea
f(\r_i, \r_j) &=& \exp[-\beta v(\r_i, \r_j)] - 1 ,\\
f^{*}(\r_i, \r_j) &=& q(\r_i, \r_j)\exp[-\beta v(\r_i, \r_j)] - 1 .
\label{deff}
\eea
Then we have from the definition of $V(i,j)$, Eqs.~(\ref{poten}) and 
(\ref{map}), that
\be
\phi(\r_i,\la_i;\r_j,\la_j) = \left\{ \begin{array}{r @{\quad 
     \mbox{if} \quad}l}
   f(\r_i,\r_j) & \la_i =\la_j \\ f^*(\r_i,\r_j) & \la_i \neq\la_j .
   \end{array} \right.    \label {dphi}
\ee

The grand canonical function is now
\be
\Xi = \sum_{N=0}^{\infty} \frac{1}{N!} \int d 1 \cdots d N 
\sum_{\la_1, \ldots, \la_N} \prod_{i=1}^N z^* (i, \la_i) \prod_{i>j}^N 
\left[ 1 + \phi(i,j) \right] .
\label{grd}
\ee
This expression is best represented diagrammatically, as
\bea
\Xi = 1 &+&
\begin{picture}(300, 20)
\put(10,3){\circle*{5}}
\put(35,0){+}
\multiput(70,3)(24,0){2}{\circle*{5}}
\put(119,0){+}
\multiput(155,3)(24,0){2}{\circle*{5}}
\thicklines
\put(155,3){\line(1,0){24}}
\put(204,0){+}
\multiput(240,-5)(24,0){2}{\circle*{5}}
\put(253,14){\circle*{5}}
\end{picture}
\nonumber \\
&+&
\begin{picture}(300, 40)
\thicklines
\multiput(10,-5)(24,0){2}{\circle*{5}}
\put(22,15){\circle*{5}}
\put(10,-5){\line(3,5){12}}
\put(66,0){+}
\multiput(104,-5)(24,0){2}{\circle*{5}}
\put(116,15){\circle*{5}}
\put(104,-5){\line(3,5){12}}
\put(117,15){\line(3,-5){12}}
\put(161,0){+}
\multiput(200,-5)(24,0){2}{\circle*{5}}
\put(212,15){\circle*{5}}
\put(200,-5){\line(1,0){24}}
\put(200,-5){\line(3,5){12}}
\put(213,15){\line(3,-5){12}}
\put(250,0){+ \,\, \ldots}
\end{picture}
\eea
In these diagrams, each circle corresponds to a function $z^*(i, \la_i)$, and 
is therefore associated with a discrete spin variable as well as with a 
position coordinate. Each line corresponds to $\phi(i,j)$. For each {\it black
circle}, we integrate over the space coordinate and sum over the spin 
coordinate. For example,
\be
\begin{picture}(35, 20)
\multiput(0,3)(24,0){2}{\circle*{5}}
\thicklines
\put(0,3){\line(1,0){24}}
\end{picture}
= \int d \r_i \, d \r_j \sum_{\la_i}\sum_{\la_j} z^*(\r_i, \la_i)\,
z^*(\r_j, \la_j) \, \phi(\r_i, \la_i; \r_j, \la_j) .
\ee
We will also have {\it white circles} over which there is neither integration 
nor summation. Note, finally, that the symmetry factor of the diagrams, being
a purely combinatorial quantity, is the same here as for the usual diagrams
without spin variables. While this is not necessarily so when the lines can
stand for different functions as they can here, nevertheless it remains true
because the identity of the lines is uniquely determined by spin variables 
which are then summed upon.

It is now easy to see that all the usual definitions and theorems which hold
for the usual spinless diagrams hold also for the spin diagrams introduced 
here. In particular, the Morita-Hiroike lemmas (see, e.g., Hansen and McDonald 
\cite{hansen}) hold here as well, provided the functional derivatives which
appear in them are generalized to be the spin-functional derivatives introduced
in the previous section. For example, we have the following generalized lemma.

{\bf Lemma}. If $\Gamma$ is a diagram consisting of black $t(\r, \la)$-circles
and $\phi$ bonds, we have that
\bea
\frac{\dd \Gamma}{\delta t(\x, \al)} &=& \mbox{\{all diagrams obtained by 
replacing a black $t$-circle of} 
\nonumber \\
& & \mbox{ $\Gamma$ by a white 1-circle labeled $(\x, \al)$\}} . \label{lem}
\eea

From another lemma of Morita and Hiroike, we have immediately that (see 
Hansen and McDonald \cite{hansen}),
\bea
\ln \Xi &=& \{\mbox{all simple connected diagrams consisting
of black $z^*$-circles and $\phi$-bonds}\} \nonumber \\
&=&
\begin{picture}(300, 20)
\put(10,3){\circle*{5}}
\put(33,0){+}
\multiput(57,3)(24,0){2}{\circle*{5}}
\thicklines
\put(57,3){\line(1,0){24}}
\put(105,0){+}
\multiput(130,-5)(24,0){2}{\circle*{5}}
\put(142,15){\circle*{5}}
\put(130,-5){\line(3,5){12}}
\put(142,15){\line(3,-5){12}}
\put(174,0){+}
\multiput(200,-5)(24,0){2}{\circle*{5}}
\put(212,15){\circle*{5}}
\put(200,-5){\line(1,0){24}}
\put(200,-5){\line(3,5){12}}
\put(212,15){\line(3,-5){12}}
\put(242,0){+ \, \ldots ,}
\end{picture}
\eea
from which we can show that
\bea
\ln \left[ \rho^{(1)}(\x, \al) / z^* (\x, \al) \right] &=& \{ \mbox{all simple
diagrams consisting of one white} \nonumber \\
& & \mbox{1-circle labeled $(\x,\al)$, one or more
black} \nonumber \\
& & \mbox{$\rho^{(1)}$-circles and $\phi$ bonds, such that they are}
\nonumber \\
&& \mbox{free of connecting circles} \} .
\eea
A connecting circle is a circle the removal of which (along with the bonds 
which emerge from it) causes the diagram to become disconnected.

From Eq.~(\ref{defc}) and the lemma (\ref{lem}), we immediately obtain that
\bea
c (\x, \al; \y, \g) &=& \{ \mbox{all simple diagrams that consist of two white}
\nonumber \\
& & \mbox{1-circles labeled $(\x,\al)$, and  $(\y,\g)$, black 
$\rho^{(1)}$-circles} \nonumber \\
& & \mbox{and $\phi$ bonds, and are free of connecting circles} \}
\label{diagc} \nonumber \\
&=& \,
\begin{picture}(300, 20)
\multiput(0,3)(24,0){2}{\circle{6}}
\thicklines
\put(3,3){\line(1,0){18}}
\put(48,0){+}
\multiput(72,-5)(24,0){2}{\circle{6}}
\put(84,15){\circle*{6}}
\put(75,-5){\line(1,0){18}}
\put(84,15){\line(-3,-5){10}}
\put(84,15){\line(3,-5){10}}
\put(120,0){+}
\multiput(144,-5)(24,0){2}{\circle{6}}
\multiput(144,15)(24,0){2}{\circle*{6}}
\put(147,-5){\line(1,0){18}}
\put(147,15){\line(1,0){18}}
\put(144,-2){\line(0,1){18}}
\put(168,-2){\line(0,1){18}}
\put(192,0){+}
\multiput(216,-5)(24,0){2}{\circle{6}}
\multiput(216,15)(24,0){2}{\circle*{6}}
\put(218,-2){\line(6,5){19}}
\put(219,15){\line(1,0){18}}
\put(218,15){\line(6,-5){20}}
\put(216,-2){\line(0,1){18}}
\put(240,-2){\line(0,1){18}}
\put(250,0){+ \, \ldots}
\end{picture}  \nonumber \\
\eea
In all these diagrams, one of the white circles is labeled $(\x, \al)$ and the
other is labeled $(\y, \g)$.

The Ornstein-Zernike relation, Eq.~(\ref{oz}), now determines $h^{(2)}$. For 
vanishing magnetic field $h = 0$, we have significant simplifications, since 
$\rho^{(1)} (\z, \la) = (1/s) \rho$, where $\rho = \langle N \rangle / V$ is 
the usual density. Furthermore, $h^{(2)}(\x, \al; \y, \g)$ and 
$c(\x, \al;\y,\g)$ now depend on $\x - \y$ provided $v(\x, \y) = v(\x - \y)$ 
and  $p(\x, \y) = p(\x - \y)$, which we always assume. 
In this case, the Ornstein-Zernike relation is very simple in Fourier 
space, i.e.,
\be
\hat{h}^{(2)}(\vec{k}, \al,\g) = \frac{\hat{c}(\vec{k}, \al,\g)}
{ 1 - (\rho/s)\hat{c}(\vec{k}, \al,\g)} ,
\label{hft}
\ee
where
\bea
\hat{h}^{(2)}(\vec{k}, \al,\g) &=& \int d \x \, e^{-i \vec{k} (\x -\y)}
\, h^{(2)}(\x, \al; \y, \g) , \nonumber \\
\hat{c}(\vec{k}, \al,\g) &=& \int d \x \, e^{-i \vec{k} (\x -\y)}
\, c(\x, \al; \y, \g) .
\eea

Going now to the percolation picture, we notice that Eq.~(\ref{S}) for $S$ can 
be rewritten
\be
S = 1 + \rho \hat{\gd} (0) .
\label{sg}
\ee
From the definition of $\gd$, Eq.~(\ref{gdag}), and from Eqs.~(\ref{ozper}), 
(\ref{cdag}) and (\ref{hft}), we have now
\be
S = \frac{1}{1 - \rho \hat{c}^{\dagger} (0)} = \frac{1}{ 1 - \rho \lim
\limits_{{s \to 1  \atop h \to 0}}\, \left[ \hat{c} (0, \al, \al) - \hat{c} 
(0, \al, \g) \right]} .
\label{sc}
\ee

This equation, together with the expansion of $c(\x,\al;\y,\g)$, 
Eq.~(\ref{diagc}), defines an expansion of $S$ in powers of the density. The 
critical density $\rho_c$ is the radius of convergence of this series, which we
can find by using standard methods of convergence analysis (see, e.g., ref. 
\cite{stanley}).

\renewcommand{\thesection}{\Roman{section}}
\section{EQUIVALENCE WITH CONIGLIO -- DeANGELIS -- FORLANI EXPANSION}
\label{seccdf}
\renewcommand{\thesection}{\arabic{section}}
\setcounter{equation}{0}

In 1977, Coniglio, DeAngelis and Forlani  (CDF) \cite{coniglio} obtained a 
density expansion for $c^{\dagger}(\x,\y)$ from a completely different point of
view. We discuss now the connection of the present work to the CDF expansion.

Coniglio {\it et al.} start by considering a classical fluid (no spins), where
the inter-particle interaction is $v(i,j)$. For this fluid, the classical 
Ornstein-Zernike direct correlation function, $c_f (\x, \y)$, has a density 
expansion given by Eq.~(\ref{diagc}), except that the white 1-circles are 
labeled  just $\x$ and $\y$ and the bonds represent the Mayer $f$-function 
defined as $f(i,j) \equiv \exp[- \beta v(i,j)] - 1$, in the same notation used
here in Eq.~(\ref{deff}). Coniglio {\it et al.} then assume that the Mayer 
$f$-function can be decomposed into a sum $f(i,j) = \fd(i,j) + 
f^* (i,j)$ where $\fd$ is related to the particles being connected and
$f^*$ to their being unconnected. The specific definitions of $\fd$ and
$f^*$ used by Coniglio {\it et al.} are relevant only to physical clustering in
a gas, but the assumption of the decomposition of $f(i,j)$ was extended by 
analogy to other systems by several workers \cite{bug,des,oth}. However, no
formal basis was provided for the definition of $\fd$ and $f^*$ in the
general case. If we take $f^*$ as formally defined here in Eq.~(\ref{deff}), 
which is natural since this function is directly related to the probability of
the two particles being unconnected, we find that
\be
\fd(i,j) \equiv f(i,j) - f^* (i,j) = p(i,j) \exp \left[ -\beta v(i,j)
\right] . \label{fdag}
\ee
This result is satisfying since $\fd$ turns out to be directly 
proportional to the probability of the two particles being connected.

Coniglio {\it et al.} then replace $f$ in the diagrammatic expansion of 
$c_f (\x,\y)$, with the sum $\fd + f^*$, and consider all the 
decomposed diagrams thus
obtained, in which every bond is either an $\fd$ or an $f^*$ function.
The sum of all the diagrams which contain at least one continuous path of 
$\fd$ functions between the white 1-circles, they consider to be 
$c^{\dagger}(\x,\y)$, while the remainder is called $c^* (\x,\y)$, so that
$c_f(\x, \y) = c^{\dagger}(\x,\y) + c^* ( \x,\y)$.

It is hardly obvious now that this algorithm agrees with the expansion obtained
here. Indeed, the function $\fd$ never appears in the present expansion. 
Nonetheless, the two formulations are equivalent. 

The proof of the equivalence holds diagram by diagram. Therefore we can take 
for simplicity a specific example, e.g., the diagram
\begin{picture}(40, 30)
\multiput(5,-5)(24,0){2}{\circle{6}}
\multiput(5,15)(24,0){2}{\circle*{6}}
\multiput(8,-5)(3,0){7}{\circle*{1}}
\multiput(8,15)(3,0){7}{\circle*{1}}
\multiput(5,-2)(0,3){7}{\circle*{1}}
\multiput(29,-2)(0,3){7}{\circle*{1}}
\end{picture}. In this diagram , every dotted line denotes a function $f(i,j)$.
Then, if we consider all the diagrams obtained by replacing every $f$-bond by 
either an $f^*$-bond or an $\fd$-bond in all possible ways, and keep only those
diagrams which contain a continuous $\fd$ path between the two white circles, 
we obtain
\be
CDF \left[ \quad
\begin{picture}(40, 20)
\multiput(5,-5)(24,0){2}{\circle{6}}
\multiput(5,15)(24,0){2}{\circle*{6}}
\multiput(8,-5)(3,0){7}{\circle*{1}}
\multiput(8,15)(3,0){7}{\circle*{1}}
\multiput(5,-2)(0,3){7}{\circle*{1}}
\multiput(29,-2)(0,3){7}{\circle*{1}}
\end{picture} \right] = 
\begin{picture}(40, 30)
\multiput(5,-5)(24,0){2}{\circle{6}}
\multiput(5,15)(24,0){2}{\circle*{6}}
\multiput(8,-5)(3,0){7}{\circle*{1}}
\put(7,14){\line(1,0){19}}
\put(7,16){\line(1,0){19}}
\put(4,-3){\line(0,1){18}}
\put(6,-3){\line(0,1){18}}
\put(28,-3){\line(0,1){18}}
\put(30,-3){\line(0,1){18}}
\end{picture} +
\begin{picture}(40, 30)
\multiput(5,-5)(24,0){2}{\circle{6}}
\multiput(5,15)(24,0){2}{\circle*{6}}
\put(7,-4){\line(1,0){19}}
\put(7,-6){\line(1,0){19}}
\multiput(8,15)(3,0){7}{\circle*{1}}
\multiput(5,-2)(0,3){7}{\circle*{1}}
\multiput(29,-2)(0,3){7}{\circle*{1}}
\end{picture},
\ee
where a double line denotes a function $\fd(i,j)$, and $CDF [ diagram ]$ on the
left hand side means
the result of the CDF algorithm when applied to the given diagram. The dotted 
lines on the right hand side ($f$-bonds) arise because once a continuous $\fd$ 
path exists, all the diagrams obtained from further replacements of $f$-bonds 
by $\fd$-bonds and $f^*$-bonds must be counted. This sum can be expressed 
more compactly by keeping the original $f$-bonds.

We now wish to prove the equivalence of this scheme to the one we have 
developed here, i.e,
\be
CDF \left[ \quad
\begin{picture}(40, 30)
\multiput(5,-5)(24,0){2}{\circle{6}}
\multiput(5,15)(24,0){2}{\circle*{6}}
\multiput(8,-5)(3,0){7}{\circle*{1}}
\multiput(8,15)(3,0){7}{\circle*{1}}
\multiput(5,-2)(0,3){7}{\circle*{1}}
\multiput(29,-2)(0,3){7}{\circle*{1}}
\end{picture} \right] = 
\lim_{s \to 1} \left[\quad
\begin{picture}(40, 30)
\multiput(5,-5)(24,0){2}{\circle{6}}
\multiput(5,15)(24,0){2}{\circle*{6}}
\thicklines
\put(8,-5){\line(1,0){18}}
\put(8,15){\line(1,0){18}}
\put(5,-2){\line(0,1){18}}
\put(29,-2){\line(0,1){18}}
\put(0, -17){$\al$}
\put(25, -17){$\al$}
\end{picture}
 - 
\begin{picture}(40, 30)
\multiput(5,-5)(24,0){2}{\circle{6}}
\multiput(5,15)(24,0){2}{\circle*{6}}
\thicklines
\put(8,-5){\line(1,0){18}}
\put(8,15){\line(1,0){18}}
\put(5,-2){\line(0,1){18}}
\put(29,-2){\line(0,1){18}}
\put(0, -17){$\al$}
\put(25, -17){$\eta$}
\end{picture}\right],
\ee
where on the right hand side, every diagram is of the spin type, and every 
full line represents a function $\phi(i,j)$. To simplify still further, let us
ignore the integration over the coordinates of the black circles (but
{\it not} the summation over spins in the spin diagrams). The integrals can 
then be added at the end of the proof. E.g., in this section we shall assume 
that
\be
\begin{picture}(40, 30)
\multiput(5,-5)(24,0){2}{\circle{6}}
\multiput(5,15)(24,0){2}{\circle*{6}}
\multiput(8,-5)(3,0){7}{\circle*{1}}
\multiput(8,15)(3,0){7}{\circle*{1}}
\multiput(5,-2)(0,3){7}{\circle*{1}}
\multiput(29,-2)(0,3){7}{\circle*{1}}
\end{picture} = 
\begin{picture}(40, 30)
\multiput(5,-5)(24,0){2}{\circle{6}}
\multiput(5,15)(24,0){2}{\circle*{6}}
\multiput(8,-5)(3,0){7}{\circle*{1}}
\multiput(8,15)(3,0){7}{\circle*{1}}
\multiput(5,-2)(0,3){7}{\circle*{1}}
\multiput(29,-2)(0,3){7}{\circle*{1}}
\put(1,-19){1}
\put(26,-19){2}
\put(1,22){3}
\put(26,22){4}
\end{picture} = f(1,3)f(3,4)f(4,2)f(1,2),
\ee
and
\be
\begin{picture}(40, 30)
\multiput(5,-5)(24,0){2}{\circle{6}}
\multiput(5,15)(24,0){2}{\circle*{6}}
\thicklines
\put(8,-5){\line(1,0){18}}
\put(8,15){\line(1,0){18}}
\put(5,-2){\line(0,1){18}}
\put(29,-2){\line(0,1){18}}
\put(-2, -19){1,$\al$}
\put(23, -19){2,$\eta$}
\put(1,22){3}
\put(26,22){4}
\end{picture} = \sum_{\la_3,\la_4}\phi(1,3)\phi(3,4)\phi(4,2)\phi(1,2),
\ee
where the labels $1,2,3,4$ denote the spatial coordinates of the vertices. Note
that in the last diagram, $\la_1 = \al$, $\la_2 = \eta$ are fixed, but we sum 
over the spins $\la_3$ and $\la_4$.

With this convention, we are ready to prove the equivalence of the two 
formulations. To do this, let us define a new {\it ``bond percolation''} 
problem on the four-vertex graph 
\begin{picture}(40, 30)
\multiput(5,-5)(24,0){2}{\circle{6}}
\multiput(5,15)(24,0){2}{\circle*{6}}
\thicklines
\put(8,-5){\line(1,0){18}}
\put(8,15){\line(1,0){18}}
\put(5,-2){\line(0,1){18}}
\put(29,-2){\line(0,1){18}}
\end{picture}. In this new problem, every bond can be either ``open'' (in which
case its two ends are said to be connected) or closed (in which case the ends
are disconnected). Let us further assume that the bond connecting the vertices
$i$ and $j$ is open with a probability $\fd (i,j)$ and closed with a 
probability $f^* (i,j)$. Formally these are not true probabilities since their 
sum is not normalized to 1, but this is of no importance. One can always 
normalize the probabilities by dividing by the proper product of $f$ functions.
With this convention, every diagram obtained from the CDF algorithm now 
represents the {\it (unnormalized) probability} of a specific configuration of
open and closed bonds on the four-vertex graph. E.g, consider
\be
\begin{picture}(40, 30)
\multiput(5,-5)(24,0){2}{\circle{6}}
\multiput(5,15)(24,0){2}{\circle*{6}}
\multiput(8,-8)(6,0){3}{$*$}
\put(7,14){\line(1,0){19}}
\put(7,16){\line(1,0){19}}
\put(4,-3){\line(0,1){18}}
\put(6,-3){\line(0,1){18}}
\put(28,-3){\line(0,1){18}}
\put(30,-3){\line(0,1){18}}
\end{picture} = \fd(1,3)\fd(3,4)\fd(4,2)f^*(1,2), \label{diagval}
\ee
where a line of $*$ represents a function $f^*$. The same diagram can also be 
thought of as a configuration of open and closed bonds on the four-vertex 
graph, by letting a double line represent an open bond and a $*$-line a closed
bond. This geometrical configuration occurs with a probability $\fd(1,3)
\fd(3,4)\fd(4,2)f^*(1,2)$, which is just the value of the equivalent diagram, 
Eq.~(\ref{diagval}). Every diagram therefore functions in a double capacity. 
On the one hand it represents an actual configuration of open and closed bonds
on some graph. On the other hand, it represents some functional value. This 
functional value is just the (unnormalized) probability of the bond 
configuration represented by the same graph.

Now, two vertices which are connected by a path of open bonds can be said to 
belong to the same cluster. In particular, the (unnormalized) probability that
the two white circles belong to the same cluster, denoted $P_{cl}(1,2)$ is the
sum of all the diagrams in which a continuous path of open bonds exists 
between these circles. This however, is exactly the class of diagrams selected
by the CDF algorithm. Hence,
\be
P_{cl}(1,2) = CDF \left[ \quad
\begin{picture}(40, 30)
\multiput(5,-5)(24,0){2}{\circle{6}}
\multiput(5,15)(24,0){2}{\circle*{6}}
\multiput(8,-5)(3,0){7}{\circle*{1}}
\multiput(8,15)(3,0){7}{\circle*{1}}
\multiput(5,-2)(0,3){7}{\circle*{1}}
\multiput(29,-2)(0,3){7}{\circle*{1}}
\end{picture} \right].
\ee
Therefore, what we now need to prove is that
\be
P_{cl}(1,2) = \lim_{s \to 1} \left[\quad
\begin{picture}(40, 30)
\multiput(5,-5)(24,0){2}{\circle{6}}
\multiput(5,15)(24,0){2}{\circle*{6}}
\thicklines
\put(8,-5){\line(1,0){18}}
\put(8,15){\line(1,0){18}}
\put(5,-2){\line(0,1){18}}
\put(29,-2){\line(0,1){18}}
\put(0, -17){$\al$}
\put(25, -17){$\al$}
\end{picture}
 - 
\begin{picture}(40, 30)
\multiput(5,-5)(24,0){2}{\circle{6}}
\multiput(5,15)(24,0){2}{\circle*{6}}
\thicklines
\put(8,-5){\line(1,0){18}}
\put(8,15){\line(1,0){18}}
\put(5,-2){\line(0,1){18}}
\put(29,-2){\line(0,1){18}}
\put(0, -17){$\al$}
\put(25, -17){$\eta$}
\end{picture}\right].
\ee
The prove this we show, by paralleling the derivation given in I, that there 
exists a mapping of the new bond percolation problem onto an $s$-spin model. 
Then $P_{cl}(1,2)$ ( which plays here the part of the pair connectedness) is 
equal to the difference of two spin diagrams. The formal proof of this claim 
is derived in the Appendix. It is easily seen to hold for every diagram in the 
CDF expansion, and shows therefore that this expansion is indeed equivalent to
the one derived here in the context of the Potts fluid.

\renewcommand{\thesection}{\Roman{section}}
\section{APPLICATION TO EXTENDED HYPERCUBES}
\renewcommand{\thesection}{\arabic{section}}
\setcounter{equation}{0}

Let us see now how the expansion obtained in section III can be applied 
to finding the critical density $\rho_c$. The system we consider consists of 
$D$ dimensional hypercubes which have an impenetrable (hard) core of side 
$a$, surrounded with a hypercubic permeable shell  of side $d > a$, so that two
particles are bound if their shells overlap. In other words, if $\r_i = 
(x_i^1, \ldots, x_i^D)$ and $\r_j = (x_j^1, \ldots, x_j^D)$ are the positions
of two particles, then
\bea
v(i,j) &=& \left\{ \begin{array}{r @{\quad} l}
            \infty & \mbox{if for all\quad} 1 \le k \le D, \quad 
            \vert x_i^k - x_j^k \vert < a \\ 
              0 &  \mbox{otherwise.}\end{array} \right. \\
              p(i,j) &=& \left\{ \begin{array}{r @{\quad} l}
             1 & \mbox{if for all \quad} 1 \le k \le D, \quad \vert x_i^k - 
             x_j^k \vert < d \\ 
             0 &  \mbox{otherwise.}\end{array} \right.
\eea
Hence, from Eq.~(\ref{deff}),
\bea
f(\r_i,\r_j) &=& \left\{ \begin{array}{r @{\quad} l}
            - 1 & \mbox{if for all \quad} 1 \le k \le D , \quad
              \vert x_i^k - x_j^k  \vert < a \\ 
              0 &  \mbox{otherwise.}\end{array} \right. \\
f^*(\r_i,\r_j) &=& \left\{ \begin{array}{r @{\quad} l}
           - 1  & \mbox{if for all \quad} 1 \le k \le D , \quad
             \vert x_i^k - x_j^k  \vert < d \\ 
             0  &  \mbox{otherwise.}\end{array} \right.
\label{fstr}
\eea
We choose this rather unrealistic system because for any dimension $D$ the 
required calculations always turn out to be much simpler for cubes than 
for the more realistic spheres. Since the aim at present is to prove the 
usefulness of 
the method rather than investigate a specific system, one should not be overly
troubled by this choice. Let us note, however, that the spherical version of
this model has been used to model microemulsions \cite{bug}, so that it is at 
least mildly relevant to some real systems. It is also the simplest model
which contains interactions, and therefore a good testing case.

Let us now define two functions of a single variable, 
\bea
F_1 (x) &=& \left\{ \begin{array}{r @ {\quad \mbox{if} \quad} l}
           1 & \vert x \vert < a \\
           0 & \vert x \vert > a .\\
           \end{array} \right. \\
F_2 (x) &=& \left\{ \begin{array}{r @ {\quad \mbox{if} \quad} l}
           1 & \vert x \vert < d \\
           0 & \vert x \vert > d .\\
           \end{array} \right. 
\eea
The equations (\ref{fstr}) can now be rewritten as
\bea
f(x^1, \ldots, x^D) &=& - \prod_{i=1}^D F_1 (x^i) .\\
f^*(x^1, \ldots, x^D) &=& - \prod_{i=1}^D F_2 (x^i) .
\eea

We shall now calculate $c^{\dagger}$ up to the second order in $\rho$ (i.e., up
to square diagrams). The zeroth order is
\be
\lim_{s \to 1} \, \left[ 
\begin{picture}(40, 10)
\multiput(5,0)(24,0){2}{\circle{6}}
\thicklines
\put(8,0){\line(1,0){18}}
\put(0, -12){$\al$}
\put(25, -12){$\al$}
\end{picture}
- \quad
\begin{picture}(40, 10)
\multiput(5,0)(24,0){2}{\circle{6}}
\thicklines
\put(8,0){\line(1,0){18}}
\put(0, -12){$\al$}
\put(25, -12){$\g$}
\end{picture} 
\right] = 
\int d \z \, \left[f(\z) - f^*(\z) \right] = d^D \left (1 -  \eta^D \right) ,
\ee
where we have defined the {\it aspect ratio}, $\eta$, as
\be
\eta \equiv \frac{a}{d} .
\ee

For the first order (triangular diagrams), we need, e.g., the integral 
\be
\Omega (\y) \equiv \int d \x \, f(\x)\, f(\x - \y) = \prod_{i=1}^D \left\{
\int d x^i \, F_1 (x^i) \, F_1 (x^i - y^i ) \right \} .\label{product}
\ee
It is precisely the factorization of this and similar integrals into a product
of identical one dimensional integrals that is the main simplification 
afforded by the use of a cubic geometry. As a beneficial side-effect, it 
allows us to check the influence of system dimensionality on the critical 
density.

$\Omega (\y)$ is very simple to evaluate when we realize that $\int d x \, 
F_1 (x) \, F_1 (x - y ) $ is merely the overlap of two segments of length $a$,
one of which is centered at the origin, the other centered at the position $y$.
By direct inspection, we have therefore that
\bea
\int d x \, F_1 (x) \, F_1 (x - y ) &=& \left\{ \begin{array}
{r @{\quad \mbox{if} \quad} l}
            2 a - \vert y \vert  &  \vert y \vert < 2 a \\         
             0  &   \vert y \vert > 2 a \\  \end{array} \right. ,
\label{f1f1} \\
\int d x \, F_2 (x) \, F_2 (x - y ) &=& \left\{ \begin{array}
{r @{ \quad \mbox{if} \quad} l}
            2 d - \vert y \vert  &  \vert y \vert < 2 d \\ 
             0  &   \vert y \vert > 2 d \\  \end{array} \right. ,
\label{f2f2} \\
\int d x \, F_1 (x) \, F_2 (x - y ) &=& \left\{ \begin{array}
{r @{ \quad \mbox{if} \quad} l}
            2a &  \vert y \vert < d - a \\
            d + a - \vert y \vert  &  d - a < \vert y \vert < d + a \\ 
             0  &   \vert y \vert > d + a  \\ \end{array} \right. .
\label{f1f2}
\eea

Let us consider a typical contribution from the triangular diagrams, e.g., the
integral $\int d x \, d y \, F_2 (x) \, F_2 (x - y ) \, F_1 (y)$. In this 
integral, we interpret the term $F_1(y)$ as merely specifying the limits of 
integration, i.e., determining that $ \vert y \vert < a$. Since $d > a$, we 
have from Eq.~(\ref{f2f2}) that
\be
\int d x \, d y \, F_2 (x) \, F_2 (x - y ) \, F_1 (y) = \int_{-a}^a dy \left[
\int d x \, F_2 (x) \, F_2 (x - y ) \right] = 4 d a - a ^2 .
\ee
All the required one dimensional integrals can now be calculated along the same
lines. In fact, equations (\ref{f1f1})--(\ref{f1f2}) suffice to calculate all 
the triangular and square diagrams, except for the last fully connected square 
diagram which is a product of six $F$ functions. For this diagram the previous
arguments have to be generalized but it can be done easily along the lines 
already presented. 

All the integrals we need are presented in Tables I--IV. The first column in 
each table contains the notation of the integral, and the third column its 
value. All these integrals are $D$-th powers of some one dimensional integrals,
[see, e.g., Eq.~(\ref{product})]. The corresponding one-dimensional integrals 
appear in the second column. 

We now have, after some straightforward calculations, that the contribution of
the triangular diagrams, denoted $\left(2 d \right)^{2D} k_2 $, is, in the 
notation of Tables I--IV,
\be
\left(2 d \right)^{2D} k_2 = I_1 - 2 \, I_2 + I_3  .
\label{k2}
\ee

Similarly, the contribution from the square diagrams is denoted $\left( 2 d 
\right)^{3D} k_3$, and in the notation of Tables I--IV is equal to
\bea
\left( 2 d \right)^{3D} k_3 &=& \frac{3}{2} J_1 - 5 J_2 + 5 J_3 - \frac{3}{2}
J_4 - 3 K_1 + 3 K_2 - \frac{1}{2} K_3 + 7 K_4 -10 K_5 + \frac{7}{2} K_6
\nonumber \\
&+& \frac{1}{2}L_1 - L_2 -L_3 + \frac{5}{2}L_4 - L_5 .\label{k3}
\eea

Substituting these results into Eq.~(\ref{sc}) yields, for the mean cluster 
size,
\be
S = \frac{1}{1 - \rho d^D \left( 1 - \eta^D \right) - \rho^2 (2 d)^{2 D} k_2
-\rho^3 (2 d )^{3D} k_3 + O(\rho^4)} .
\label{sers}
\ee

The critical density is commonly measured in dimensionless units. Let us define
\be
B = ( 2 d )^{D} \rho ,   \label{defb}
\ee
which is chosen to reduce to the total excluded volume when $a \to 0$ 
\cite{bc}. As mentioned in the introduction, this reduces the dependence of the
density on the details of the system and facilitates the presentation of the 
results. Finally, using standard methods \cite{abra}, we find the series 
expansion for the mean cluster size to be
\be
S = 1 +S_1 B + S_2 B^2 + S_3 B^3 + O(B^4) , \label{sercl}
\ee
with
\bea
S_1 &=& \frac{1 - \eta^D}{2^D} , \nonumber \\
S_2 &=& k_2 + S_1^2 , \nonumber \\
S_3 &=& k_3 + 2 S_1 k_2 + S_1^3 . 
\eea

\renewcommand{\thesection}{\Roman{section}}
\section{RESULTS FOR HYPERCUBES}
\renewcommand{\thesection}{\arabic{section}}
\setcounter{equation}{0}

To find the critical density, we need some extrapolation method which will 
yield the point of divergence of the series (\ref{sercl}). Unfortunately, we
know very few terms in this series. The results can be improved, however, by 
using additional information. In particular, $S$ diverges as $S \sim \left(B_c
- B\right)^{-\g}$, and $\g$ appears to be universal, i.e., independent of the 
interaction, and furthermore, equal to the well known value it takes in 
lattice percolation. {\it Biased methods} use the known value of $\g$ to 
calculate $B_c$. There are several methods available, but for the present use,
the best seems to be the one developed by Arteca, Fernandez, and Castro 
(AFC) \cite{afc}. The biased version of this method 
relies on the fact that the function $(B_c-B)^{\g} S(B)$ is analytical around 
$B_c$. Therefore, we define a function of two variables,
\be
g(u,B) = \left(1 - \frac{B}{u} \right)^{\g} S(B) , \label{defgu}
\ee
where we have assumed a known value for $\g$. Given an expansion $S = \sum S_n 
B^n$, we can expand $g(u,B)$ in powers of $B$, with the result,
\be
g(u,B) = \sum_{n=0}^{\infty} g_n(u) B^n ,
\ee
where
\be
g_n(u) = \sum_{k=0}^n {\g \choose k} (- u )^{-k} S_{n-k} .
\ee
Since $g(u,B)$ is analytical at $u = B_c$, the idea is now to look for values 
of $u$ which maximize the convergence of the series 
$\{g_n(u)\}_{n=1}^{\infty}$ as $n \to \infty$. The AFC proposal is to generate
a series $\{u_n\}_{n=1}^{\infty}$ of solutions of the equation
\be
g_n(u_n) = 0 . \label{geq}
\ee
Then we expect that
\be
\lim_{n \to \infty} \, u_n = B_c .
\ee
In the case of the series (\ref{sercl}), we know the terms only up to $n=3$,
which is too little to extrapolate a reliable limit to the series 
$\{u_n\}_{n=1}^{\infty}$. Therefore, we take simply the last term of the series
, $u_3$, as our estimate of $B_c$.

In figures (1)--(5), we compare this theoretical estimate to results of 
computer simulations for dimensions 2, 3, 4 and 5, as a function of the aspect
ratio $\eta$. The simulations were performed with an efficient new algorithm we
introduced recently for investigating continuum percolation with interactions 
\cite{algo}. Unlike the usual Metropolis algorithm, this method increases
serially the density in the system until percolation is achieved. The effect of
the interactions is included through a rejection criterion which produces an
effective statistical weight for the configurations. The final (percolating) 
configuration contained close to 30,000 particles in two and three dimensions 
and around 10,000 in the higher dimensions. The results are averages over 10 
independent runs and the numerical error is estimated to range from 5\% to 
10 \%.

Let us first consider the well reproduced qualitative behavior of 
$B_c(\eta)$. The main features can be understood from simple arguments. As 
pointed out by Bug {\it et al.} \cite{bug}, the minimum of $B_c(\eta)$ results 
from the competition between two processes. As the diameter of the hard core 
increases, it becomes harder to bring particles close enough to each other for 
them to bind. This effect clearly dominates at high $\eta$. At low $\eta$, on
the other hand, the hard core is much too small relative to the soft shell to
prevent binding significantly, but it increases the average distance between 
bound particles. As a result the clusters are ``longer'', and therefore 
percolate ``sooner'', hence $B_c$ is lower.

The dependence on $D$ is understandable in terms of the ratio of the permeable 
shell volume to the hard core volume. Clearly, this ratio increases with the 
dimensionality. Therefore, at any given $\eta$, the influence of the hard core
diminishes as $D$ increases. As a result, the graph of $B_c(\eta)$ looks 
flatter, while the influence of the final singularity at $\eta = 1$ (at which 
there is no possibility of binding anymore and therefore no percolating phase),
is limited to regions of higher $\eta$. This influences also the feasibility
of the simulations. Clearly, the larger the hard core, i.e., the greater 
$\eta$, the harder it is to simulate the system. Hence, e.g., in two 
dimensions, we have no results beyond $\eta = 0.8$ due to this difficulty. 
As $D$ increases, however, higher $\eta$ become more accessible. Unfortunately,
it is generally harder to simulate high dimensional systems, which is why we 
stopped at $D=5$. The trends, however, are obvious, and there should be nothing
exceptional for $D>5$.

Turning now to the detailed comparison between theory and simulations, we see 
that the results are quantitatively close to each other. Even in the worse 
case, in two dimensions (Fig. 1), the difference between theory and 
simulations is at most 11\% (at $\eta = 0.5$). The position of the minimum, 
theoretically predicted to be $\eta = 0.79$, also agrees well with the 
simulations.

Very recently, however, Okazaki {\it et al.} \cite{okazaki} have performed new
simulations of two dimensional continuum percolation, and claim that the 
critical exponents are in fact different from those of lattice percolation. For
the exponent $\gamma$ they find $\gamma = 1.94$ instead of the higher $43/18$ 
obtained in lattice percolation \cite{stau}. More work is certainly needed to
resolve this issue. Here we merely note the effects of such a shift in the 
value of $\gamma$ on the theoretical prediction. Figure 2 presents the results
obtained in two dimensions when the value of $\gamma$ found by Okazaki 
{\it et al.} is used for the biasing. The fit is quite improved by this change.
Unfortunately the approximations we use are too drastic to claim that our 
calculations actually support the value obtained by Okazaki {\it et al.}. But
they do suggest that the question of the universality class of continuum 
percolation is not yet settled and requires more investigation.

Note however that the relative change in $B_c$ is smaller than the change in 
$\gamma$ which produced it. Therefore, $B_c$ is not extremely sensitive to the 
value of the bias. Since other simulations find values of $\gamma$ in higher
dimensions which are at least close to the values obtained in lattice 
percolation, we may use these lattice values for the bias and expect that $B_c$
will not be affected notably. For this reason and for lack of other data on the
value of $\gamma$ in continuum percolation, we therefore use the accepted 
lattice values for dimensions higher than two.

In three dimensions, up to $\eta = 0.8$, the {\it greatest} discrepancy between
theory and ``experiment'' is only 4 \%, well within the limits of the 
simulation error. Even at $\eta = 0.8$, the largest discrepancy we have, the
deviation is only about 6 \%, still within the simulation error (which 
increases slightly with $\eta$ because of the above mentioned problems in 
performing the simulations in the vicinity of $\eta=1$). 

This trend continues in dimension 4, where the discrepancy between the theory 
and the simulations remains below 6 \% until $\eta = 0.65$, and in dimension 5.
Note, however, that as a general rule simulation errors increase with 
dimensionality due to the greater difficulty in performing the simulations, and
therefore comparison with the theory becomes slightly more problematic. In 
particular, at $D= 5$ there seems to be a systematic discrepancy with the 
theory which could well reflect some finite size scaling effects in the 
simulations. It is therefore quite possible that the theory is actually more 
precise at this point than the simulations.

Considering that we use the series for $S$ only up to second order, the
quantitative agreement obtained is a remarkable success for the theory.

\renewcommand{\thesection}{\Roman{section}}
\section{DISCUSSION}
\renewcommand{\thesection}{\arabic{section}}
\setcounter{equation}{0}

We have seen how the Potts fluid mapping allows us to derive a general 
expansion in powers of the density for the mean cluster size and the pair
connectedness. This expansion can be used to calculate the percolation
threshold by calculating the first terms in the series then using some 
extrapolation method to find the radius of convergence of the series. We have
applied this scheme to interacting hypercubes in dimensions ranging from 2 
to 5, and we have shown that even with merely three terms in the series we 
already obtain very good quantitative results, within a few percents of the 
results of the computer simulations. 

Ours is not the first work to use density expansions of the mean cluster size,
obtained in one way or another (e.g., from an analogy with lattice systems 
\cite{haan}), but it is the first one to derive quantitatively adequate results
for interacting systems. We believe this success stems from a simple but
fundamental reason.

When Coniglio {\it et al.} \cite{coniglio} obtained their expansion of $S$, 
they did
not use it directly to calculate the critical density. Instead, they turned for
inspiration to the theory of liquids and considered the integral equations for 
the pair correlation which had proved successful there, primarily the Percus-
Yevick (PY) equation. Coniglio {\it et al.} derived a percolative analog of 
this equation by applying the CDF algorithm to its diagrammatic expansion. Some
time later, De Simone {\it et al.} \cite{des} applied this equation to a system
of extended
spheres, the spherical analog of the system investigated in the present work.
The results were qualitatively correct but quantitatively poor. Discrepancies
between theory and simulations ranged from 40 \% to no less than 10 \%. Several
researchers \cite{oth} extended this work to other interactions or shapes with
always the same quantitatively inadequate results. We believe this failure 
holds a simple (in hindsight, in fact, obvious), but apparently 
under-appreciated lesson.

The guiding principle behind the work of Coniglio {\it et al.} and their 
followers seems to have been viewing continuum percolation as a high density 
phenomenon.
This is very reasonable from a point of view which starts from the usual theory
of fluids and then extracts the percolative quantities from their normal fluid
analogs. The CDF algorithm is explicitly based on this point of view; given,
e.g., the direct correlation function $c_f (\r)$ of the fluid, the percolating
part, $c^{\dagger} (\r)$ can be extracted from it by looking at a diagrammatic
expansion. This means we need to know $c_f(\r)$ first. The percolation 
transition occurs at relatively high densities, so that the normal fluid is
either a dense gas or a liquid. In this case, integral equations are the main
tool for calculating the fluid's properties, and therefore an adequate starting
point for obtaining the percolative properties as well.

But such a point of view misses the most important aspect of the percolation
transition, namely that it {\it is} a phase transition. Integral equations are
known to describe the liquid well only away from the liquid-gas critical point.
As we approach the transition, they fail. The quantitative failure of the 
percolative integral equations suggests that they suffer from an analogous 
defect. There is direct evidence for this claim. Seaton and Glandt 
\cite{seaton} checked the mean cluster size for a wide range of densities and 
compared the predictions of integral equations with the simulations. They found
that the integral equations are indeed very successful away from the critical 
density, but worsen steadily as $\rho \to \rho_c$. In short, the percolative 
integral equations behave with respect to the percolation transition exactly as
their analog in liquids behave with respect to the liquid transition.

At first sight this seems puzzling. At $\rho_c$, the normal liquid is usually
away from its own critical point, and it {\it is} therefore well described by
integral equations. Now the procedure for extracting the percolative part from
these equations involves {\it no further approximations}. Why then, do we 
obtain a relatively bad approximation by extracting {\it exactly} the 
percolative part from a very {\it good} approximation?

This is precisely the heart of the matter, which is the trivial observation 
that the phase transition is entirely ruled by the {\it singular} part of the
relevant functions. Integral equations tend to smooth out this singular part, 
as evidenced, for example, by their systematic overestimation of the critical
density \cite{des}. This doesn't matter for the description of the liquid, 
because the singular part is small at these densities. However when we turn to 
the percolative analog, the entire physics we are after lies precisely in this
overly smoothed out singular part.

To understand a singular behavior, we must use methods adapted to singular
functions. This is the simple but under-appreciated lesson from all the
preceding. It requires us to abandon the view of continuum percolation as 
primarily a high density process. The density at which the transition takes 
place is irrelevant for the choice of descriptive tools, since it only 
influences the {\it analytical} part of the function. The proper tools to 
describe the percolation transition are not those which describe well 
high density systems, but rather those which describe well other phase 
transitions.

The present work proceeds clearly from this point of view. The underlying
normal liquid plays no part in the description presented here. Nowhere does the
function $c_f(\r)$ or any of its parents appear. Instead, we have a mapping 
onto a Potts fluid, from which we calculate all the relevant quantities. One
important difference between this and the older point of view is that the
magnetic transition of the Potts fluid can be {\it identified} with the 
percolation transition, since the two relevant order parameter are essentially 
the same. Therefore we are forced from the outset to employ only methods which
are useful for the description of the magnetic phase transition. Indeed, 
integral equations are not a natural tool to use here since they will clearly
fail in the description of the Potts fluid. Instead, we rely on one of the 
best tried methods in critical phenomena, namely, an expansion valid in the
disordered phase (high temperature usually, low density in the present case).
A low density expansion makes no sense from the point of view of percolation as
a high-density phenomenon. Indeed, we do not expect the actual value of $S$ 
obtained from the truncated series to be even remotely good. But what we look 
for is a failure of convergence, the appearance of a singularity, and for this
a truncated series can be singularly well adapted, as countless investigations
of the Ising model have shown repeatedly \cite{stanley}.

It is precisely because the mapping onto a Potts fluid identifies one phase
transition (percolation) with another (magnetic) that we can achieve a
remarkable {\it quantitative} agreement between theory and simulations. 
Clearly, the truncated series expansion coupled with an extrapolation method to
find the radius of convergence is at present the best available method to 
calculate theoretically the critical density in continuum percolation.

\section*{appendix}
\renewcommand{\theequation}{A\arabic{equation}}
\setcounter{equation}{0}

We consider some diagram in the expansion, e.g., the example used in section 
\ref{seccdf}, 
\begin{picture}(40, 30)
\multiput(5,-5)(24,0){2}{\circle{6}}
\multiput(5,15)(24,0){2}{\circle*{6}}
\thicklines
\put(8,-5){\line(1,0){18}}
\put(8,15){\line(1,0){18}}
\put(5,-2){\line(0,1){18}}
\put(29,-2){\line(0,1){18}}
\end{picture}. On this diagram we define a bond-percolation problem. We can now
map the problem onto a spin model by simply assigning to every vertex an 
$s$-spin, and assigning to every bond between the vertices $i$ and $j$ a 
function $\phi(i,j)$ defined as in Eq.~(\ref{defphi}). We now proceed to show, 
just as in paper I, that the bond-percolation model corresponds quantitatively
to the spin model. The result is again a geometrical mapping according to which
every percolation configuration corresponds to several spin configurations, in 
each of which all the spins belonging to a single cluster are parallel. 
Different clusters, however, have randomly assigned spin values. The following
derivation parallels closely the one presented in paper I for the continuum 
case.

Let the chosen diagram contain $n$ vertices, and let us denote 
\be
\Gamma(1,2,\ldots,n) = \prod_{\mbox{all existing} \atop {\mbox{bonds in} \atop 
\mbox{the graph}}} \!\!\!\!\!\!\! \phi(i,j).
\ee
By convention, the two white circles are always labeled $1$ and $2$. The value 
associated with the diagram in the spin model is therefore 
$\sum_{\la_3, \la_4, \ldots, \la_n} \Gamma(1,\ldots,n)$, where we assume that
the spins of the two white circles, $\la_1$ and $\la_2$, are fixed , and where
we ignore the integration as in section \ref{seccdf}. The integration can be 
restored at the very end of the proof without changing anything. For example, 
for the four-vertex graph mentioned above, we have
\be
\Gamma(1,2,3,4) = \phi(1,3)\phi(3,4)\phi(4,2)\phi(2,1).
\ee

Let us now select two spins, say $\la_i$ and $\la_j$ such that the bond 
$\phi(i,j)$ exists in the graph. In the expression for $\Gamma(1, \ldots, n)$, 
let us separate all possible configurations $\{\la_m\}$ into those where 
$\la_i = \la_j$ and the rest. Then,
\be
\sum_{\la_m}\Gamma(1,\ldots,n) = f(i,j) \sum_{\left\{{\la_m \atop  \la_i = 
\la_j}\right\}} \frac{\Gamma(1,\ldots,n)}{\phi(i,j)} + f^*(i,j) 
\sum_{\left\{{\la_m \atop  \la_i \ne \la_j}\right\}} \frac{\Gamma(1,\ldots,n)}
{\phi(i,j)}.\label{st1}
\ee
We can rewrite the sum over the spins when $\la_i \ne \la_j$ as the difference 
between the sum over the spins without constraints and the sum over the spins 
when $\la_i = \la_j$, so that
\be
\sum_{\la_m}\Gamma(1,\ldots,n) = \fd(i,j) \sum_{\left\{{\la_m 
\atop  \la_i = \la_j}\right\}}\frac{\Gamma(1,\ldots,n)}{\phi(i,j)} + f^*(i,j)
\sum_{\{\la_m\}} \frac{\Gamma(1,\ldots,n)}{\phi(i,j)} .   \label{st3}
\ee
where we used the relation $\fd (i,j) = f(i,j)- f^*(i,j)$. The last sum 
$\sum_{\{\la_m\}}$ on the right hand side is now performed over all spin 
configurations without constraints. Let us now choose another pair of spins, 
say $\la_i$ and $\la_k$ (if such a bond exists). Repeating the previous 
procedure, we obtain that
\bea
\sum_{\la_m}\Gamma(1,\ldots,n) &=& \fd(i,j)\, \fd(i,k)\sum_{\left\{{\la_m 
\atop \la_i = \la_j =\la_k}\right\}} \frac{\Gamma(1,\ldots,n)}{\phi(i,j)
\phi(i,k)} + \fd(i,j)\, f^*(i,k) \sum_{\left\{{\la_m}\atop {\la_i =\la_j}
\right\}} \frac{\Gamma(1,\ldots,n)}{\phi(i,j)\phi(i,k)}\nonumber \\
& +& f^*(i,j) \, \fd(i,k) \sum_{\left\{{\la_m} \atop {\la_i = \la_k}\right\}} 
\frac{\Gamma(1,\ldots,n)}{\phi(i,j)\phi(i,k)} + f^*(i,j)\, f^*(i,k) 
\sum_{\{\la_m\}} \frac{\Gamma(1,\ldots,n)}{\phi(i,j) \phi(i,k)}. \label{st5}
\eea

This argument can be easily generalized to all pairs of spins. Let us 
consider one of the sums into which the function $\sum_{\la_m}
\Gamma(1,\ldots,n)$ has been decomposed a step before, and consider a pair 
$(m,n)$ such that the bond $\phi(m,n)$ exists in the chosen diagram. 
Two possibilities arise:
\begin{enumerate}
\item Previous constraints already determine that $\la_m = \la_n$ (for example,
there could be some $p$ for which $\la_m = \la_p$ and $\la_n = \la_p$). Then, 
\bea
\sum_{\left\{\la_m \atop {\mbox{previous} \atop \mbox{ constraints}}\right\}}
\frac{\Gamma(1,\ldots,n)}{\phi(i,j)\cdots \phi(l,k)} &=& \fd(m,n)
\sum_{\left\{\la_m \atop {\mbox{prev.} \atop \mbox{constr.}}\right\}}
\frac{\Gamma(1,\ldots,n)}{\phi(i,j)\cdots \phi(l,k)\phi(m,n)}\nonumber \\
&+& f^*(m,n)\sum_{\left \{\la_m \atop {\mbox{prev.} \atop \mbox{constr.}}
\right\}}\frac{\Gamma(1,\ldots,n)}{\phi(i,j)\cdots \phi(l,k)\phi(m,n)} ,
\label{gen1}
\eea
where we have used the fact that if $\la_m = \la_n$, then $\phi(m,n) = 
\fd(m,n) + f^*(m,n)$.

\item Previous constraints do not determine that $\la_m = \la_n$. Then the 
situation is as it was for the pair $(i,j)$, and the sum will split in the 
following way
\bea
\sum_{\left\{\la_m \atop \mbox{constr.}\right\}}\frac{\Gamma(1,\ldots,n)}
{\phi(i,j)\cdots \phi(l,k)}&=&  \fd(m,n)\sum_{\left\{\la_m :\, \ldots \atop 
\la_m =\la_n \right\}}\frac{\Gamma(1,\ldots,n)}{\phi(i,j)\cdots \phi(l,k)
\phi(m,n)}\nonumber \\
&+ & f^*(m,n)\sum_{\left\{\la_m :\, \dots\right\}} 
\frac{\Gamma(1,\ldots,n)}{\phi(i,j)\cdots \phi(l,k)\phi(m,n)},
\label{gen2}
\eea
where in the second term on the right hand side no new constraint has been 
introduced.
\end{enumerate}

The geometrical mapping follows because in case (2), every time a factor $\fd$
appears, we have a new constraint forcing the two spins to be parallel. On the
other hand, in the percolation model, such a factor implies that the two 
vertices are connected, and therefore belong to the same cluster. When a factor
$f^*$ appears, no new constraint is added, so that the two vertices can be
assigned spins at random. In case (1), on the other hand, the two spins are
already forced to be parallel by virtue of some previous constraints. According
to the foregoing argument, this means that they are both connected (at least
indirectly) to some other spin and therefore that they already belong to the
same cluster. In this case it does not matter anymore whether these vertices 
are also linked directly (a factor $\fd$) or not (a factor $f^*$). Hence, we 
see that vertices which belong to the same cluster in the percolation picture 
must be assigned parallel spins while the clusters themselves have randomly 
assigned spin values.

When all pairs have been covered, the set of constraints of a particular sum
specifies exactly which particles belongs to which clusters in the original
bond-percolation model configuration. Since the expression for $\sum_{\la_m}
\Gamma(1,\ldots,n)$ contains sums over all possible constraints, it can be 
rewritten as a sum over all possible clusterings of the original percolation 
configuration. Thus, let us define
\be
P(\mbox{ conn.}) \equiv \prod_{\left\{{{\mbox{ all} \atop \mbox{bound}}
\atop {\mbox{pairs } \atop (i,j)}} \right\}} \!\!\! \! \fd(i,j)
\prod_{\left\{{\mbox{ all} \atop {\mbox{unbound} \atop {\mbox{pairs} \atop
(m,n)}}} \right\}} \!\!\!\! f^*(m,n).
\label{definp}
\ee
Then we can write that
\be
\sum_{\la_m}\Gamma(1,\ldots,n) = \sum_{\left\{ {\mbox{ all possible} \atop 
{\mbox{ connectivity} \atop \mbox{ states}}}\right\}}\sum_{\left\{{{\la_m 
\mbox{ consistent} \atop\mbox{ with the }} \atop {\mbox {connectivity} \atop 
\mbox{state}}}\right\}} P(\mbox{ conn.})
\label{gen3}
\ee
where the sum over all spins is consistent with the connectivity state in the
sense of the geometrical mapping, i.e., that all vertices within a single 
cluster must be assigned the same spin.

If we now assume that two vertices are connected with an unnormalized 
probability $\fd$ and disconnected with an unnormalized probability $f^*$ (see
section \ref{seccdf}), we have the interpretation that $P$(conn.) is the 
unnormalized probability of finding a specific percolation configuration of 
connected and disconnected vertices. Therefore, given a function 
$F(1,\ldots,n)$  defined in the percolation model, of the coordinates of the 
vertices, we can define its ``average'' as
\be
\left\langle F (1, \ldots, n) \right\rangle_p = \sum_{\left\{{\mbox{conn.} 
\atop \mbox{states}}\right\}}P(\mbox{conn.}) \> F (1,\ldots,n).
\label{rel}
\ee
Similarly, we can have averages performed in the Potts fluid, which will be
denoted by $\left\langle{\; }\right\rangle_s$. Given a quantity $G$ defined in 
the Potts fluid, we have
\be
\left\langle G (1,\la_1; \ldots; n, \la_n) \right\rangle_s = 
\sum_{\{\la_m \}} G (1,\la_1; \ldots; n, \la_n) \prod_{i>j}\phi(i,j).  
\label{pav}
\ee
Note that $\la_1$ and $\la_2$ are assumed fixed. Eq.~(\ref{gen3}) now implies 
that in general, for any quantity $G (1,\la_1; \ldots; n, \la_n)$ defined in 
the Potts fluid system, we have that
\be
\left\langle G (1,\la_1; \ldots; n, \la_n) \right\rangle_s = 
\left\langle  \sum_{\{\la_m \mbox{: cl}\}} G (1,\la_1; \ldots; n, \la_n)
\right\rangle_p ,\label{fundam}
\ee
where $\sum_{\{\la_m \mbox{: cl}\}}$ means a summation over all spin 
configurations which are consistent with a given clustering in the sense of the
geometrical mapping (i.e., such that all the spins in a single cluster are 
parallel). 

This fundamental relation allows us to relate to each other quantities in the 
percolation model and quantities in the spin model. In particular, let us 
consider the function $\Omega (i,j)$  defined in the percolation model as
\be
\Omega (i,j)= \left\{ \begin{array}{r @{\qquad}l}
   1 & \mbox{if \quad}i , j \mbox{\quad belong to the same cluster}\\
   0 & \mbox{otherwise} 
   \end{array} \right. .   \label {omega}
\ee
If we now denote $P_{cl} (1,2)$ the probability that the two white circles 
labeled $1$ and $2$ belong to the same cluster, then
\be
P_{cl}(1,2) = \langle \Omega(1,2) \rangle_p . \label{pcon}
\ee

We shall now prove that
\be
\left\langle \Omega(1,2) \right\rangle_p = \lim_{s \to 1} \left\langle 
\frac{1}{s-1} \sum_{\la_1, \la_2} \p(\la_1)\p(\la_2) \right\rangle_s,
\ee
where $\p(\la)$ equals $(s-1)$ if $\la = 1$ and $-1$ otherwise [see 
Eq.~(\ref{psi})].

By using the fundamental relation  Eq.~(\ref{fundam}), we have
\be
\left\langle \frac{1}{s-1} \sum_{\la_1, \la_2}\p(\la_1)\p (\la_2)
\right\rangle_s =  \left\langle \frac{1}{s-1} \sum_{\{\la_m:\, cl.\}} 
\sum_{\la_1, \la_2} \p(\la_1)\p (\la_2)\right\rangle_p 
\ee
We now separate the average on the right hand side into two parts: when 
$\la_1$ and $\la_2$ belong to the same cluster and when they do not.
Then, from the definition of the percolation average, Eq.~(\ref{rel}),
\bea
\left\langle \sum_{\{\la_m:\, cl.\}} \sum_{\la_1, \la_2}\p(\la_1)\p (\la_2)
\right\rangle_p &=& \sum_{\left\{{\mbox{conn.} \atop \mbox{states}}\right\}}P 
(\mbox{conn.})\sum_{\{\la_m:\, cl.\}} \sum_{\la_1, \la_2}\p(\la_1)\p (\la_2)
\Omega (1,2) \nonumber \\
&+& \sum_{\left\{{\mbox{conn.} \atop \mbox{states}}\right\}} P(\mbox{conn.}) 
\sum_{\{\la_m:\, cl.\}} \sum_{\la_1, \la_2}\p(\la_1)\p (\la_2) 
\left[ 1 - \Omega (1,2)\right].\label{omeg2}
\eea
The first sum contributes only if $\la_1, \la_2$ belong to the same cluster, 
while the second contributes only if they belong to separate clusters. Because
of the geometrical mapping, $\p(\la_1)= \p (\la_2)$ in the first sum. Also, in
the sum over $\{ \la_m: \, cl\}$, every cluster other than the one containing
$\la_1$ and $\la_2$ contributes a factor $s$, the number of possible spin
assignments. The cluster containing $\la_1$ and $\la_2$ , on the other hand, 
contributes a factor $(s-1)^2$ if $\la_1=\la_2=1$ and $(-1)^2$ for the $(s-1)$ 
other possible choices for $\la_1 = \la_2$. Therefore,
\be
\sum_{\{\la_m:\, cl.\}} \sum_{\la_1, \la_2}\p(\la_1)\p (\la_2)\Omega (1,2) = 
s^{K -1}\left[(s-1)^2 + (s-1)(-1)^2 \right]\Omega (1,2)
\ee
where $K$ is the total number of clusters in the configuration. Hence,
\bea
\sum_{\left\{{\mbox{conn.} \atop \mbox{states}}\right\}} 
\frac{P(\mbox{conn.})}{s-1}
\sum_{\{\la_m:\, cl.\}} \sum_{\la_1, \la_2}\p(\la_1)\p (\la_2) \Omega (1,2) &=&
\sum_{\left\{{\mbox{conn.}\atop \mbox{states}}\right\}}P(\mbox{conn.})s^{K -1}
(s-1+1)\Omega (1,2)\nonumber \\
&=& \left\langle s^{K}\Omega(\la_1,\la_2)\right\rangle_p . \label{omeg}
\eea

The second term on the right hand side of Eq.~(\ref{omeg2}) contributes only 
if $\la_1, \la_2$ belong to different clusters. The cluster containing $\la_1$ 
contributes a factor $(s-1)$ if $\la_1 = 1$, and a factor $(-1)$ in all the 
other $(s-1)$ cases. The same holds for the cluster containing $\la_2$. The 
$K - 2$ remaining clusters contribute each a factor $s$. Since the values of 
$\la_1$ and $\la_2$ are assigned independently of each other, we have that
\be
\sum_{\{\la_m:\, cl.\}} \sum_{\la_1, \la_2} \p(\la_1)\p (\la_2)
\left[ 1 - \Omega (1,2)\right]= s^{K -2}\left[(s-1) + (s-1)(-1)\right]^2 
\left[1 - \Omega (1,2)\right] = 0 . \label{omegnot}
\ee

Combining Eq.~(\ref{omeg}) and (\ref{omegnot}), we see that
\be
\lim_{s\to 1} \left\langle \frac{1}{s-1} \sum_{\la_1, \la_2} \p(\la_1)\p(\la_2)
\right\rangle_s = \left\langle \Omega (1,2)\right\rangle_p . \label{lemnew}
\ee

Now, separating the case  $\la_1 = \la_2 \equiv \sigma$ from the case 
$\la_1 \equiv \sigma \ne \la_2 \equiv \xi$, we have that
\be
\left\langle \sum_{\la_1, \la_2} \p(\la_1)\p(\la_2)
\right\rangle_s = \sum_{\sigma}\p^2(\sigma)R(\sigma,\sigma) + \sum_{\sigma 
\ne \xi}\p(\sigma)\p(\xi)R(\sigma,\xi),
\ee
where 
\be
R(\la_1,\la_2) \equiv \sum_{\la_3,\ldots,\la_n}\Gamma(\la_1,\la_2, \la_3,
\ldots,n), \label{defR1}
\ee
in which we have written explicitly the spin variables in the function 
$\Gamma(1,2,\ldots,n)$ for clarity. It is important to see that because of the
summation over $\la_3,\ldots, \la_n$, $R(\la_1,\la_2)$ is independent of the 
specific value of $\la_1$ and $\la_2$, and depends only on whether they are 
equal. Because of this, we now have that
\be
\sum_{\sigma}\p ^2(\sigma)R(\sigma,\sigma) = (s-1)^2
R( \al=1, \al=1) + (s-1) (-1)^2 R(\sigma=\al, \sigma=\al) ,
\ee
where $\al$ is some arbitrary value of the spin different from $1$. The 
factor $(s-1)$ in the last term on the right hand side represents the possible
choices of this spin $\al \ne 1$. As a result,
\be
\lim_{s \to 1} \, \frac{1}{s-1}\sum_{\sigma}\p ^2(\sigma) \,
R(\sigma, \sigma) = \lim_{s \to 1} \,R(\al, \al) 
\label{term1}
\ee
where $\al\ne 1$, but is otherwise arbitrary. 

Similarly,
\be
\sum_{\sigma \ne \xi}\p(\sigma)\p (\xi) \, R(\sigma, \xi) = (s-1)^2 \left[ 
R(1,\al) + R(\al, 1) \right] + (s-1)(s-2) \, R(\al, \eta),
\ee
where $\al \ne\eta$ are arbitrary values of the spin which are both different
from $1$. As a result,
\be
\lim_{s \to 1} \, \frac{1}{s-1}\sum_{\sigma\ne \xi}\p(\sigma)\p (\xi)\,
R(\sigma, \xi) = - \lim_{s \to 1}\,  R(\al, \eta).
\label{term2}
\ee
Combining Eqs.~(\ref{term1}) and (\ref{term2}) we have that
\be
\lim_{s\to 1} \left\langle \frac{1}{s-1} \sum_{\la_1, \la_2} \p(\la_1)\p(\la_2)
\right\rangle_s  = \lim_{s \to 1} \left[ R(\al, \al) - R(\al, \eta) \right].
\label{alm}
\ee
where $\al, \eta \ne 1$ and $\al \ne \eta$. Comparing with Eqs.~(\ref{pcon}),
(\ref{lemnew}) and (\ref{defR1}), we finally obtain that
\be
P_{cl}(1,2) = \lim_{s \to 1} \left[  \sum_{\la_3,\ldots,\la_n} 
\Gamma(\al,\al, \la_3,\ldots,\la_n) - \sum_{\la_3,\ldots,\la_n} 
\Gamma(\al,\eta, \la_3,\ldots,\la_n) \right].
\ee
For the specific example of the four-vertex graph mentioned at the beginning of
this section, for example, this means that
\be
P_{cl}(1,2) =  CDF \left[ \quad
\begin{picture}(40, 30)
\multiput(5,-5)(24,0){2}{\circle{6}}
\multiput(5,15)(24,0){2}{\circle*{6}}
\multiput(8,-5)(3,0){7}{\circle*{1}}
\multiput(8,15)(3,0){7}{\circle*{1}}
\multiput(5,-2)(0,3){7}{\circle*{1}}
\multiput(29,-2)(0,3){7}{\circle*{1}}
\end{picture} \right] = 
\lim_{s \to 1} \left[\quad
\begin{picture}(40, 30)
\multiput(5,-5)(24,0){2}{\circle{6}}
\multiput(5,15)(24,0){2}{\circle*{6}}
\thicklines
\put(8,-5){\line(1,0){18}}
\put(8,15){\line(1,0){18}}
\put(5,-2){\line(0,1){18}}
\put(29,-2){\line(0,1){18}}
\put(0, -17){$\al$}
\put(25, -17){$\al$}
\end{picture}
 - 
\begin{picture}(40, 30)
\multiput(5,-5)(24,0){2}{\circle{6}}
\multiput(5,15)(24,0){2}{\circle*{6}}
\thicklines
\put(8,-5){\line(1,0){18}}
\put(8,15){\line(1,0){18}}
\put(5,-2){\line(0,1){18}}
\put(29,-2){\line(0,1){18}}
\put(0, -17){$\al$}
\put(25, -17){$\eta$}
\end{picture}\right].
\ee

This clearly holds for all the diagrams in the CDF expansion. If we restore the
integration on the black circles, we find that this proves the equivalence,
diagram by diagram, of the CDF algorithm and the expansion we derived in the 
present paper.

\begin{table}[b]
\caption{The integrals contributed by triangular diagrams. The first
column is the notation and the last is the value. This value is the $D$--th 
power of the one dimensional integral whose integrand occupies the second 
column.The variables $x$ and $y$ are integrated upon.}
\begin{center}
\begin{tabular}{c c c}
$I_1$  & $F_1(x)\,F_1(x-y)\,F_1(y)$ & $( 3 a^2)^D $\\ 
$I_2$  & $F_2(x)\,F_2(x-y)\,F_1(y)$ & $(4d a - a^2 )^D$ \\
$I_3$  & $F_2(x)\,F_2(x-y)\,F_2(y)$ & $(3 d ^2)^D $\\ 
\end{tabular}
\end{center}
\end{table}

\begin{table}
\caption{The integrals contributed by square diagrams with four lines.
The first column is the notation and the last is the value. This value is the 
$D$--th power of the one dimensional integral whose integrand occupies the 
second column. The variables $x$, $y$, and $z$ are integrated upon.}
\begin{center}
\begin{tabular}{c c c} 
$J_1$  & $F_1(x)\,F_1(x-y)\,F_1(z)\,F_1(z-y)$ & $\left(\frac {16}{3}a ^3
\right)^D$ \\
$J_2$  & $F_1(x)\,F_1(x-y)\,F_2(z)\,F_2(z-y)$ & $\left( 8 d a^2 - \frac{1}{3} 
a^3\right)^D$ \\ 
$J_3$  & $F_1(x)\,F_2(x-y)\,F_2(z)\,F_2(z-y)$ & $\left( 6 d^2 a - \frac{2}{3} 
a ^3\right)^D $\\ 
$J_4$  & $F_2(x)\,F_2(x-y)\,F_2(z)\,F_2(z-y)$ & $\left(\frac {16}{3}d ^3
\right)^D$ \\ 
\end{tabular}
\end{center}
\end{table}

\begin{table}
\caption{The integrals contributed by square diagrams with five lines.
The first column is the notation and the last is the value. This value is the 
$D$--th power of the one dimensional integral whose integrand occupies the 
second column. The variables $x$, $y$, and $z$ are integrated upon.}
\begin{center}
\begin{tabular}{c c c} 

$K_1$  & $F_1 (y) \,F_1(x)\,F_1(x-y)\,F_1(z)\,F_1(z-y)$ & $\left( \frac{14}{3} 
a^3\right)^D$ \\ 
$K_2$  & $F_1 (y) \,F_1(x)\,F_1(x-y)\,F_2(z)\,F_2(z-y)$ & $\left( 6 d a^2 -
\frac{4}{3} a ^3 \right)^D$ \\ 
$K_3$  & $F_1 (y) \,F_2(x)\,F_2(x-y)\,F_2(z)\,F_2(z-y)$ & $\left( 8 d^2 a - 4 d
a^2 + \frac{2}{3} a ^3 \right) ^3 $\\ 
$K_4$  & $F_2 (y) \,F_1(x)\,F_2(x-y)\,F_1(z)\,F_2(z-y)$ & $\left(8 d a^2 - 
\frac{10}{3}a^3 \right)^D$ \\ 
$K_5$  & $F_2 (y) \,F_1(x)\,F_2(x-y)\,F_2(z)\,F_2(z-y)$ & $\left(6 d^2 a - d 
a^2 - \frac{1}{3} a^3 \right)^D$ \\ 
$K_6$  & $F_2 (y) \,F_2(x)\,F_2(x-y)\,F_2(z)\,F_2(z-y)$ & $\left( \frac{14}{3}
 d^3\right)^D$ \\ 
\end{tabular}
\end{center}
\end{table}

\begin{table}
\caption{The integrals contributed by square diagrams with six lines.
The first column is the notation and the last is the value. This value is the 
$D$--th power of the one dimensional integral whose integrand occupies the 
second column. The variables $x$, $y$, and $z$ are integrated upon.}
\begin{center}
\begin{tabular}{c c c} 
$L_1$  & $F_1 (x)\,F_1(y)\,F_1 (z) \,F_1(x-y)\,F_1(x - z)\,F_1(z-y)$ & 
$\left (4 a^3 \right)^D$ \\ 
$L_2$  & $F_2 (x)\,F_2(y)\,F_2 (z) \,F_1(x-y)\,F_1(x - z)\,F_1(z-y)$ & 
$\left(6 d a^2 - 2 a^3 \right) ^D$ \\ 
$L_3$  & $F_1 (x)\,F_2(y)\,F_2 (z) \,F_2(x-y)\,F_2(x - z)\,F_1(z-y)$ & 
$\left(8 d a^2 - 4 a^3 \right)^D$ \\ 
$L_4$  & $F_2 (x)\,F_2(y)\,F_2 (z) \,F_2(x-y)\,F_2(x - z)\,F_1(z-y)$ & 
$\left(6 d^2 a - 2 d a^2 \right)^D$ \\ 
$L_5$  & $F_2 (x)\,F_2(y)\,F_2 (z) \,F_2(x-y)\,F_2(x - z)\,F_2(z-y)$ & 
$\left (4 d^3 \right)^D$ \\ 
\end{tabular}
\end{center}
\end{table}

\begin{figure}
\begin{center}
\leavevmode
\epsfysize=500pt
\epsffile{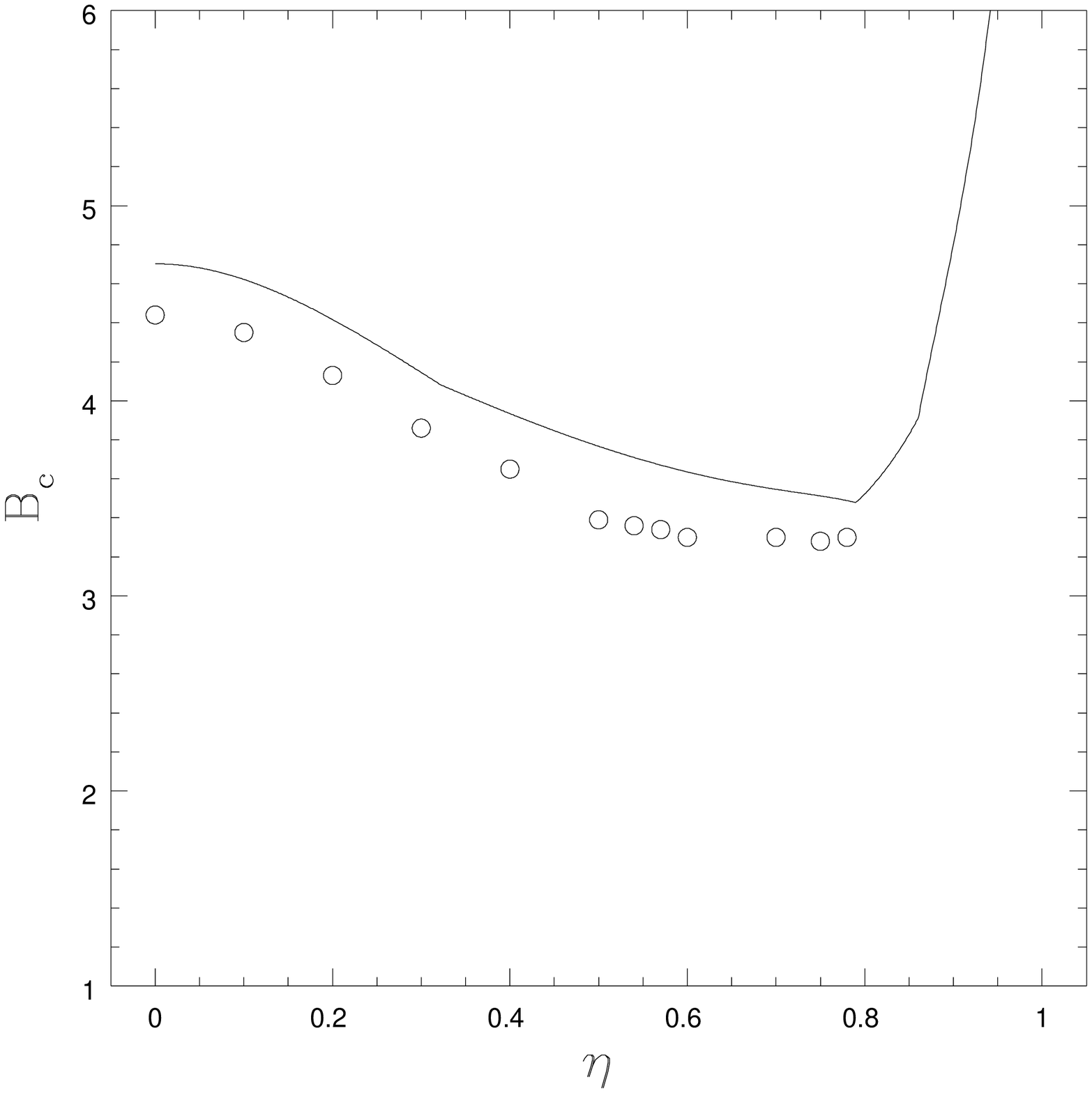}
\end{center}
\caption{The percolation threshold, $B_c$, as a function of the aspect 
ratio for hard core squares with a soft shell, in dimension 2. The line 
represents the theoretical calculation, obtained from the AFC algorithm, 
biased with a value of $\gamma = 43/18$ [19]. The circles are the 
results of computer simulations.}
\end{figure}

\begin{figure}
\begin{center}
\leavevmode
\epsfysize=500pt
\epsffile{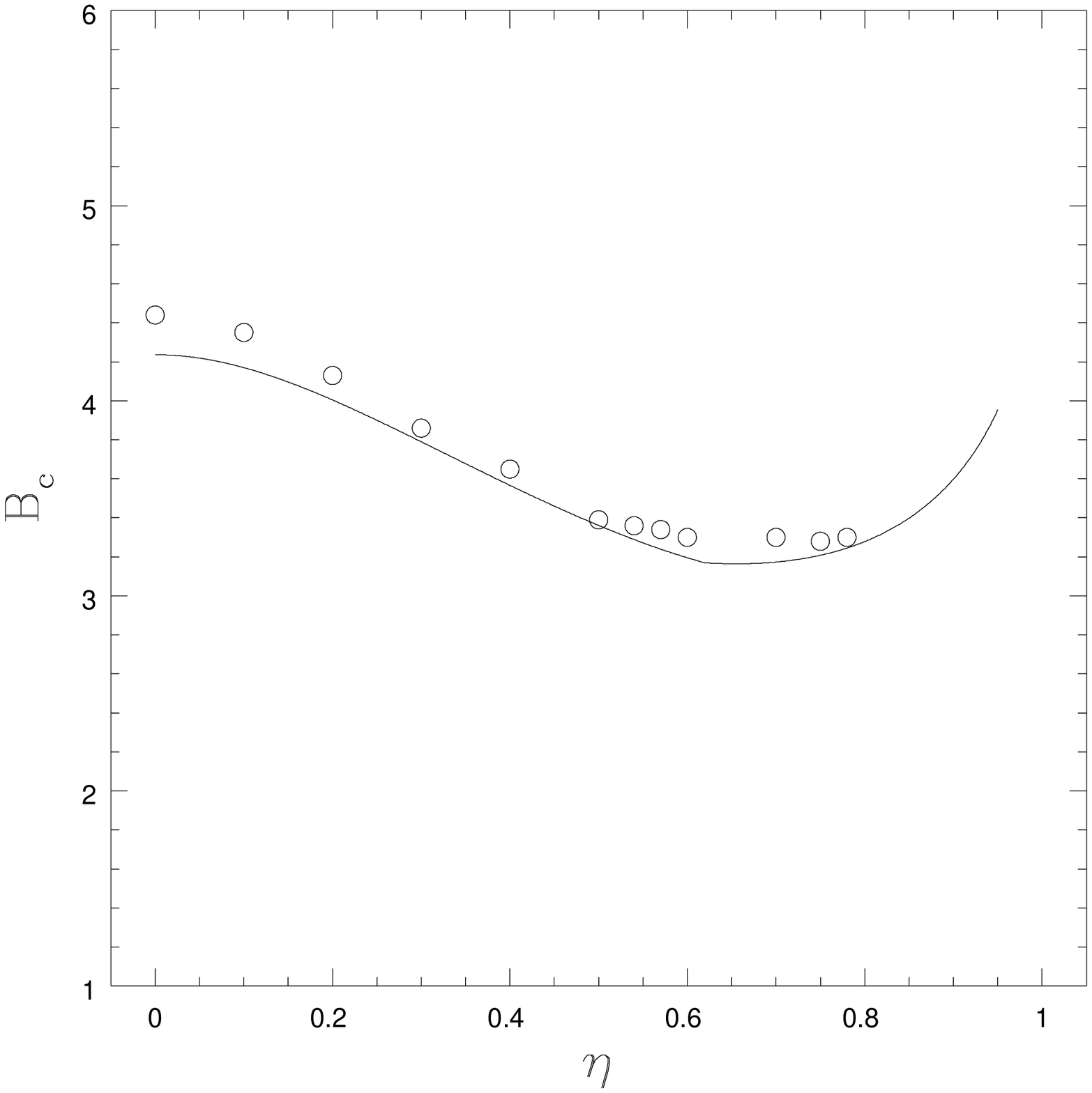}
\end{center}
\caption{The percolation threshold, $B_c$, as a function of the aspect 
ratio for hard core squares with a soft shell, in dimension 2. The line 
represents the theoretical calculation, obtained from the AFC algorithm, 
biased with a recently obtained value of $\gamma = 1.94$ [6]. The circles are
the results of computer simulations.}
\end{figure}

\begin{figure}
\begin{center}
\leavevmode
\epsfysize=500pt
\epsffile{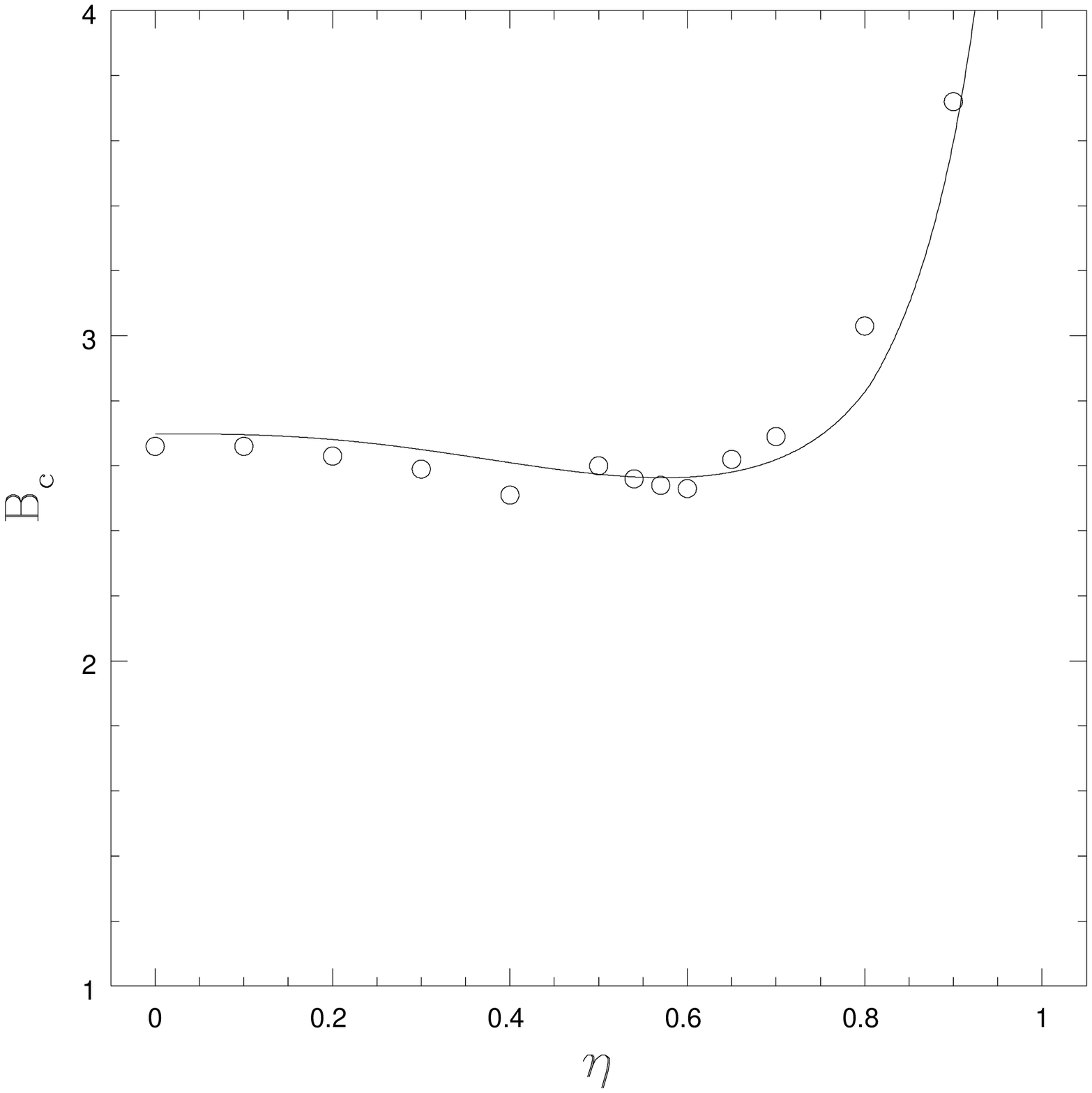}
\end{center}
\caption{The percolation threshold, $B_c$, as a function of the aspect 
ratio for hard core cubes with a soft shell, in dimension 3. The line 
represents the theoretical calculation, obtained from the AFC algorithm, biased
with a value of $\g = 1.74$ [20]. The circles are the results of computer
simulations.}
\end{figure}

\begin{figure}
\begin{center}
\leavevmode
\epsfysize=500pt
\epsffile{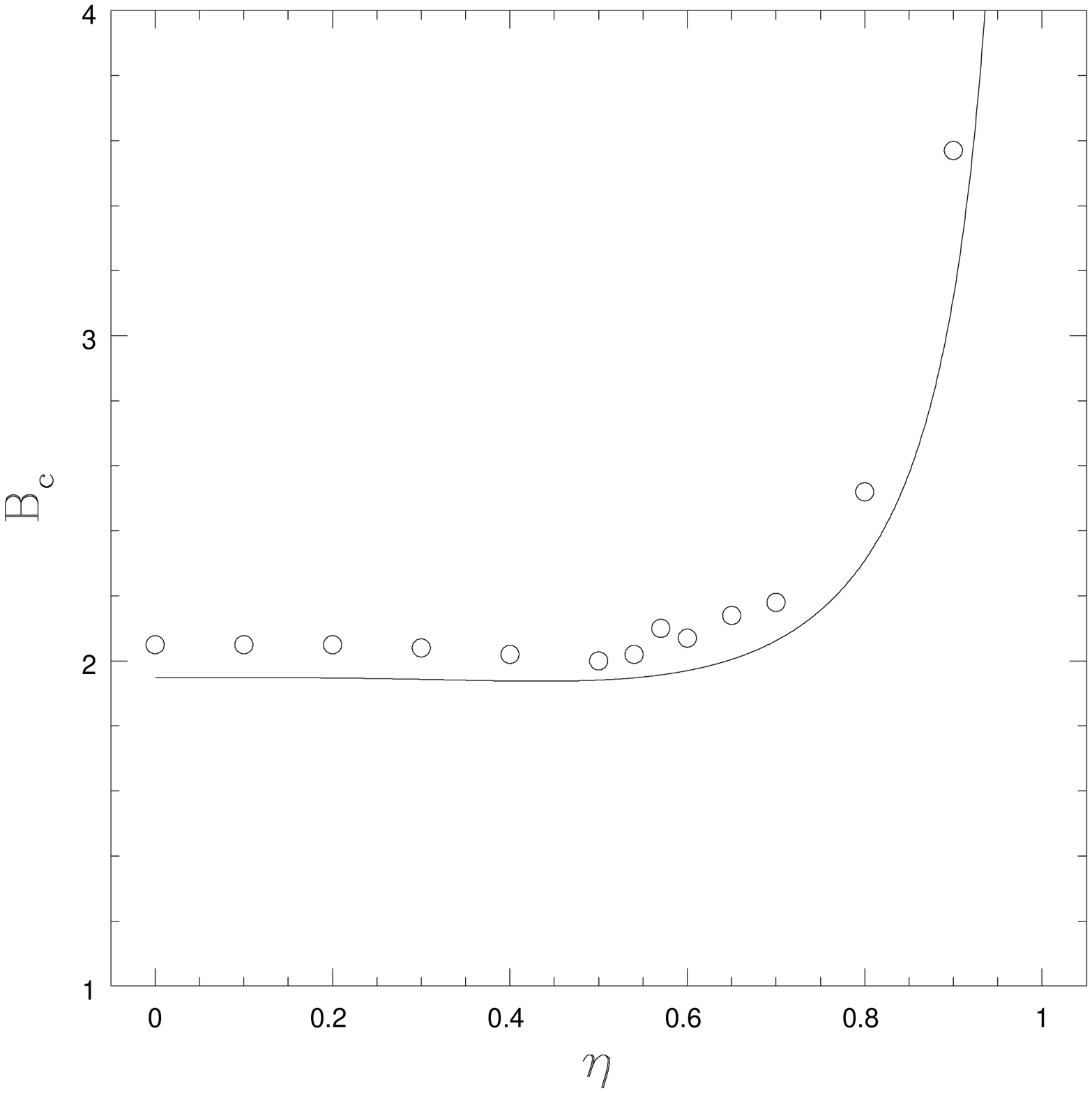}
\end{center}
\caption{The percolation threshold, $B_c$, as a function of the aspect 
ratio for hard core hypercubes with a soft shell in dimension 4. The line 
represents the theoretical calculation, obtained from the AFC algorithm, biased
with a value of $\g = 1.44$ [19]. The circles are the results of 
computer simulations.}
\end{figure}

\begin{figure}
\begin{center}
\leavevmode
\epsfysize=500pt
\epsffile{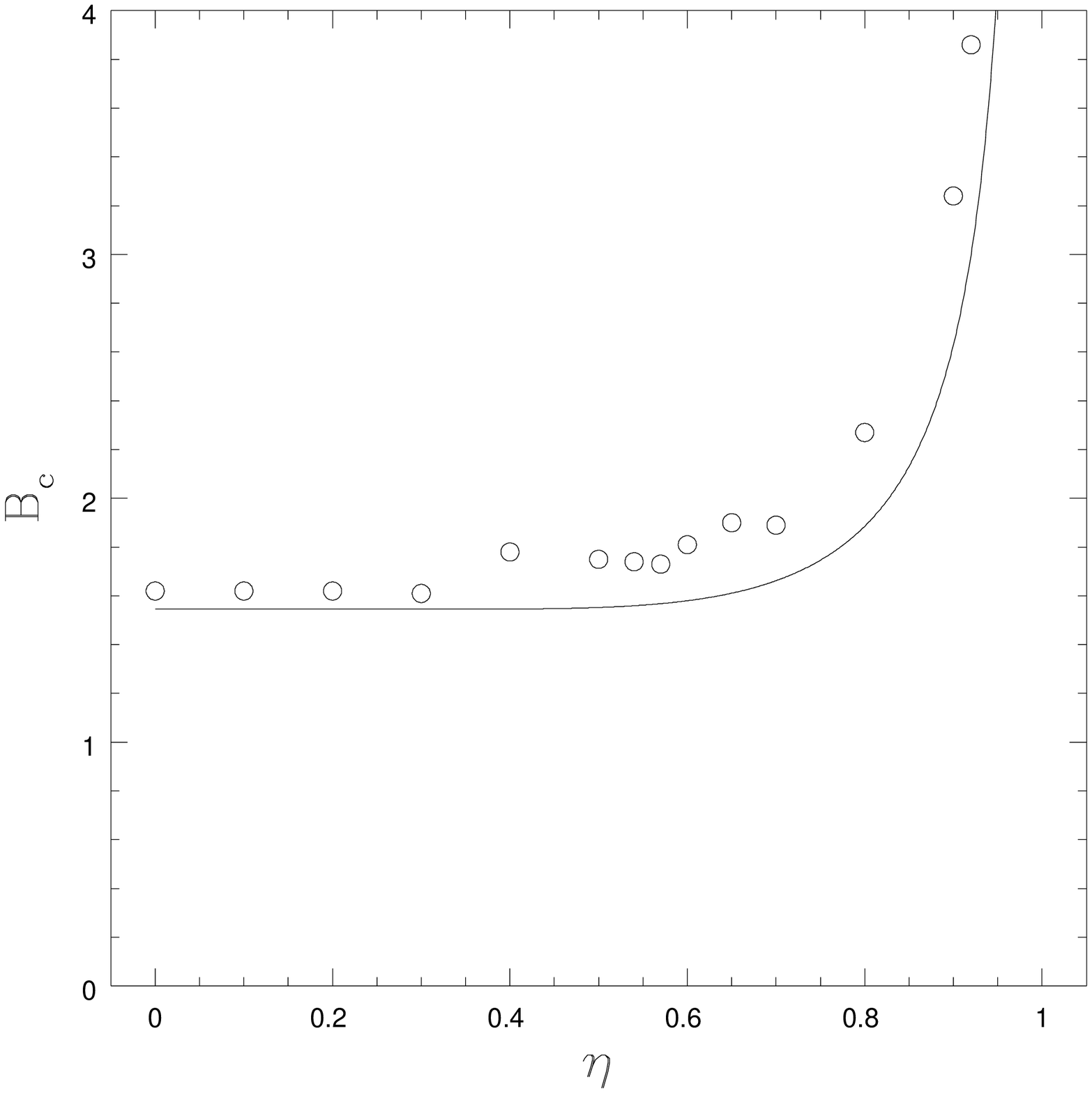}
\end{center}
\caption{The percolation threshold, $B_c$, as a function of the aspect 
ratio for hard core hypercubes with a soft shell in dimension 5. The line 
represents the theoretical calculation, obtained from the AFC algorithm, biased
with a value of $\g = 1.2$ [19]. The circles are the results of computer
simulations.}
\end{figure}

\end{document}